%% file: SearchSUSY.tex
\documentclass[twocolumn]{revtex4}

\usepackage{amsmath}
\usepackage{graphicx}
\usepackage{float}

\begin{document}

\input{definitions}

\title{Cosmo-Particle Searches for Supersymmetry at the Collider Detector at Fermilab}

\author{David Toback}
\affiliation{Department of Physics, Texas A\&M University\\
College Station, TX 77843-4242\\
toback@tamu.edu}

\date{\today}

\begin{abstract}
Some theories of particle physics are so compelling that it is worth doing a comprehensive and systematic set of experimental searches to see if they are realized in nature. Supersymmetry is one such theory. This review focuses on the motivation for a broad set of cosmology-inspired search strategies at the Tevatron and on their implementation and results at the Collider Detector at Fermilab (CDF) with the first few fb$^{-1}$ of integrated luminosity of data.
\end{abstract}

\maketitle

\newcounter{myctr}

\section{Introduction}

\mysect{Not Your Typical Introduction}

Several theories are candidates for describing particle physics which goes beyond the Standard Model (SM). Some of them are so compelling that it is worth doing a comprehensive and systematic set of searches to see if they are realized in nature. Supersymmetry (SUSY) is one such theory~\cite{SUSYOverview}. In this paper we touch on some of the theoretical motivations for SUSY, and concentrate on important possible connections to dark matter, astronomy, and cosmology. If these connections are realized in nature then cosmology-inspired models, and their implications, will help us in deciding where to focus the bulk of our experimental search efforts to discover SUSY. This paper describes a broad set of cosmo-particle searches for SUSY at the Tevatron using the Collider Detector at Fermilab (CDF)~\cite{DetectorStuff}. A review covering comparable searches at the D\O~detector (CDF's sister detector~\cite{DzeroDetector}) can be found in Ref.~\cite{Arnaud}. These searches also have important implications for when the Large Hadron Collider (LHC) at CERN becomes operational.

\subsection{Supersymmetry and What Makes it Theoretically Compelling}

There are many descriptions of SUSY and its motivations and we refer the reader to excellent reviews~\cite{SUSYOverview}. We quickly summarize the salient issues for our purposes. Essentially, SUSY is a theory that predicts a symmetry between fermions and bosons. The Minimal Supersymmetric Standard Model (MSSM) describes a scenario in which all the quarks and leptons have bosonic counterparts, denoted as squarks and sleptons, $\widetilde{q}$ and $\widetilde{l}$ respectively. Likewise, the gauge bosons, including the as-yet undiscovered Higgs, have fermionic counterparts, which are, as a group, known as the gauginos. The ElectroWeak (EWK) eigenstates of the gauginos mix and their mass eigenstates are referred to as charginos and neutralinos, for the charged and neutral supersymmetric gauge particles respectively. We typically order them via their mass hierarchy: \none, \ntwo, \nthree, \nfour~and \conepm, and \ctwopm.

\mysect{SUSY Helps in Many Ways}

SUSY helps solve a number of problems in the SM. For example, the SM predicts a divergent value for corrections to the Higgs boson's mass, but Supersymmetry offers a way around this problem~\cite{higgscorrection}. When the massive sparticles are included in the loops, having bosonic and fermionic counterparts make the value of the corrections to the mass mostly cancel, leaving behind a finite value if the sparticle masses are at or below the TeV scale. Another exciting possibility is that SUSY provides a way for the different forces' coupling constants to unify at the Grand Unified Theory (GUT) scale, provided that the sparticles that have masses below the Fermi scale. There is no {\it a priori} requirement that this happen, but the unification of the EWK and Strong forces at higher energies is very suggestive~\cite{unification}.

\mysect{Advantages and Disadvantages of SUSY}

One thing that makes SUSY difficult to look for experimentally is that it has many different versions with vastly different phenomenological features. In particular, as in the SM, the particle masses and coupling are not specified from first principles. The most model independent search method is to pick a set of trial masses and couplings and search for each version of the sparticles that way. This is not a practical approach experimentally, though, as general SUSY models have 100 free parameters~\cite{SUSYOverview}. A better avenue incorporates theoretical principles and experimental constraints, and there are some good general phenomenological studies in this vein, for example see~\cite{hewriz}. One of the most important constraints is that since SM particles can decay though other particles in loops, any version of SUSY which predicts or allows a short proton lifetime ($<10^{31-33}$ years~\cite{pdg}), must be incorrect. Also, any model which predicts a particle's mass to be equal to the mass of their corresponding sparticle, as is in the relationship between matter and antimatter, must be wrong; if sparticles had masses equal to their particle partners, we would have discovered bosonic electrons long ago~\cite{pdg}. For this reason, it has been argued that SUSY is a broken symmetry.

\mysect{Different Ways to Proceed}

Identifying the correct methods of symmetry breaking, or uniquely predicting the masses and/or couplings has proved elusive. One commonly suggested way to protect the lifetime of the proton, and solve other problems, is the assertion of the conservation of $R$-parity~\cite{rparity}, defined as $R=(-1)^{3(B-L)+2S}$, where $B$ is the baryon number, $L$ is the lepton number, and $S$ is the spin. This gives $R = 1$ for SM model particles and $R = -1$ for MSSM particles. This protects the proton's lifetime because there are no lighter SM particles for it to decay to that do not violate other symmetries or conservation laws. $R$-parity violating terms would have to be small for lepton number violation and still allow neutrino mixing (lepton flavor violation)~\cite{RefsfromRPV}. A side benefit of $R$-parity conservation is that it also provides an exciting possibility that SUSY could be easily and directly tied to dark matter and cosmology. 

\subsection{The Dark Matter in the Early Universe and Today}

\mysect{SUSY: a Dark Matter Candidate}

\subsubsection{Astronomy and Dark Matter}

\mysect{Astronomy: Galaxy Rotation}

To briefly review the ``dark matter problem,'' we note that models of galaxy rotation with stars and hydrogen gas have long been known to significantly underpredict the rotation velocity of stars at large radii from the center of a galaxy~\cite{GalaxyDarkMatter}. Simulations that include lots of other massive particles, whose dominant interaction is gravitational, do a much better job of matching the observed results much more closely. Since these massive particles aren't seen, i.e., the particles don't interact with photons or other SM matter much, they are called ``dark matter.'' A particle solution would also explain why most of the mass is clumped at the center of the galaxy (due to gravity), but can also form a ``halo'' around the entire galaxy as would be expected if dark matter were a particle that has only small coupling which we can't see, such as a SUSY particle.

\mysect{D. M. as Particles: Colliding Galaxies}

\begin{figure}
\figurehelper{\includegraphics[width=8.66cm]{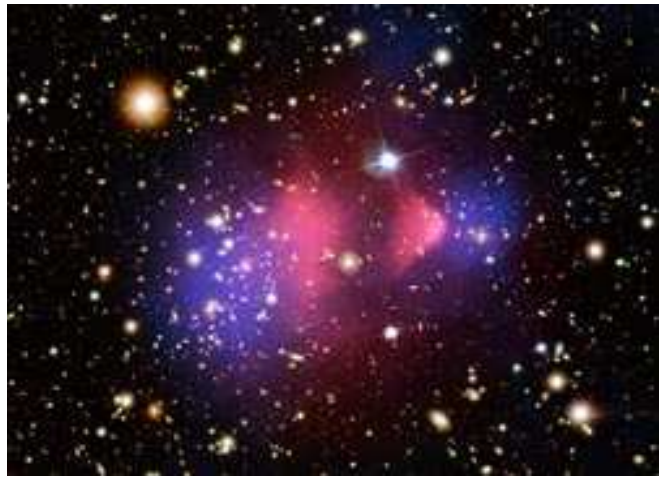}}
\caption{An image of the dark matter and baryonic mass distributions after the collision between two clusters of galaxies. The blue represents the amount of matter present, as measured by gravitational lensing, but not otherwise seen. This has been attributed to dark matter. The red represents the mass distribution, as measured by X-rays from the galaxy remnants, mostly hydrogen gas. The interpretation of this data is that the dark matter has largely passed through unaffected by the collision, but the baryonic matter has significantly slowed due to SM interactions. Taken from \protect{\cite{BGPic}}.}
\label{fig: BulletGalaxies}
\end{figure}

A second easily understood piece of evidence that the dark matter in galaxies is a particle comes from observations of collisions of galaxy 
clusters~\cite{DarkMatterCluster}. Figure~\ref{fig: BulletGalaxies} shows the collision remnants where the mass of the baryonic and dark matter components are shown separately from X-ray and gravitational lensing methods respectively. The two cluster components are well separated after the collision. The particle physics explanation is that the dark matter particles move through each other since they rarely interact, while the baryonic matter slows down significantly, due to SM interactions.

\subsubsection{Dark Matter, Particle Physics, and Cosmology}

\mysect{Astronomy Problem: Particle Solution}

Many versions of SUSY predict massive, neutral, stable particles that could explain the data. For example, if $R$-Parity is conserved then, like the proton or the electron, the lightest SUSY particle (LSP) cannot decay due to conservation of energy. While many models of new physics predict dark matter candidates one of the things that makes SUSY special is that it provides a well studied model of particle physics that can also make quantitative predictions for astronomy and cosmology. This includes predictions about early universe physics, including the baryon asymmetry as well as a prediction for the dark matter relic density imprinted in the cosmic microwave background radiation.

\mysect{Entering the Era of Precision Cosmology}

Cosmology measurements have entered a precision era. WMAP and other experiments currently have SM matter accounting for about 4\% of the mass of the known universe, and dark matter accounting for about 23\%; the remaining 73\% of the universe is called dark energy~\cite{WMAP}. While an analysis of dark energy is outside the scope of this review, the dark matter relic density, and the ratio of the mass scale of the dark matter and its coupling with ordinary matter, provides strong constraints on SUSY if indeed the lightest SUSY particle is the dark matter. If the dark matter is made of sparticles, there is the exciting possibility that SUSY not only provides a dark matter candidate, but also describes early universe physics and can provide full calculation of the dark matter relic density, $\Omega _{\textrm{SUSY dm}}$~\cite{SUSYomega}.

\subsection{Overview: A Number of Different Types of Cosmo-Particle Solutions}

\begin{figure}
\figurehelper{\includegraphics[width=8.66cm]{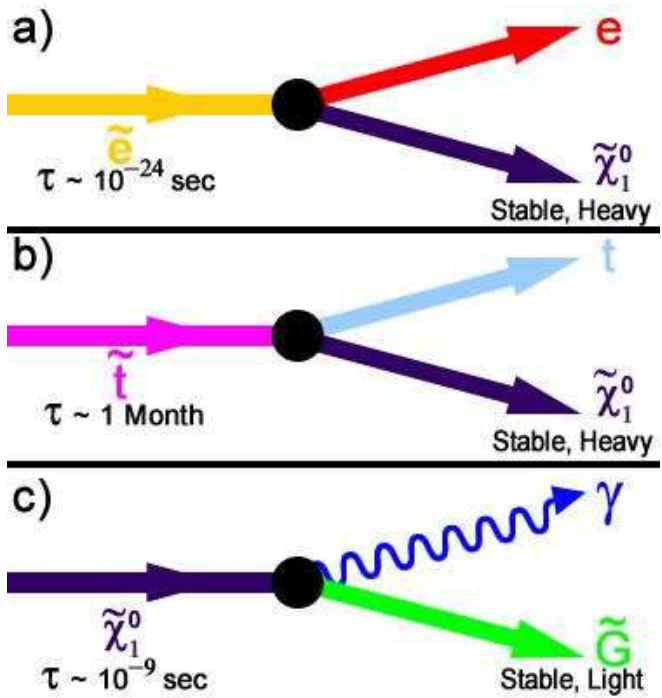}}
\caption{Different decays of SUSY particles to the LSP in $R$-parity conserving scenarios with different lifetimes ($\tau$). (a) shows the fast decay of a selectron to the lightest neutralino, which is a cold dark matter candidate in mSUGRA models. (b) shows the decay of a long-lived stop, a CHAMP in this case, to the lightest neutralino. (c) shows the decay of the lightest neutralino to a photon and a light graviton, which is a warm dark matter candidate in GMSB models. Note the wide variation of time scales through which these processes occur. This has important implications for both the early universe physics and the dark matter content of the universe today.}

\label{fig: FeynmanDecays}
\end{figure}

If nature has chosen the simple solution that 23\% of the mass of the universe is a single type of particle that is, in fact, the lightest SUSY particle, then this places very severe constraints on the types of SUSY that can produce this scenario. That being said, there are still many different types of solutions possible, even with this assumption. We next consider some possibilities, summarized in Figure~\ref{fig: FeynmanDecays}, that describe some of the crucial elements of how the early universe might have evolved into the universe we know today, and how sparticles, and their properties, fit in.

\subsubsection{Minimal Solution with Cold Dark Matter}

A minimal scenario is that the lightest sparticle is neutral, weakly interacting, stable, and mostly produced in the early universe ($t \cong 10^{-24}$~sec) as the decay product of all the heavier, short-lived sparticles (see Figure~\ref{fig: FeynmanDecays}a). It is still here today since it cannot decay and has such small interaction cross sections that it is rarely annihilated. The model is minimal in that all the dynamics are mostly over after inflation~\cite{Inflation}. Cold dark matter is favored by most cosmological models~\cite{darkmatter}, and many SUSY models have a lightest particle that fits this description in the 100~GeV/$c^2$ mass range, typically the lightest neutralino, \none~\cite{SUSYOverview}. The minimal model of SUSY that also includes potential grand unification with gravity is known as Minimal Supergravity, or mSUGRA for short~\cite{mSUGRA}. This model has become popular in the field for these reasons, and because it is well described by only a few parameters, described in more detail in Section~\ref{mSUGRA}, we refer to it as our baseline, minimal search model.

\mysect{Cold D. M.: Non-Minimal Solution}

\subsubsection{Non-Minimal Solution with Cold Dark Matter}

While the minimal models have the advantage of elegance, there is no compelling reason to believe that nature has chosen such a simple
solution. For example, it would imply that the amount of dark matter, which is much more prominent in the universe than SM particles, has
simply evolved along with the universe since $10^{-9}$ seconds after the Big Bang, when the temperature dropped below levels where the
other sparticles were typically produced. Normal matter, especially quarks and atoms, have very different roles in the universe over time.
There are many non-minimal models that could produce the dark matter we observe today. For example, there are models with long-lived
CHArged Massive ParticleS (CHAMPS) that decay into dark matter (see Figure~\ref{fig: FeynmanDecays}b)~\cite{{CHAMPTheory}}. In this case
the dominant SUSY particle of the universe on the timescale of universal inflation is a CHAMP, but it decays into the LSP/dark matter
after a nanosecond, a second, or even a month. The advantage to these models are that they essentially give an extra ``knob'' which helps in fitting models to the data. These models have recently gained favor in parts of the theoretical community for other reasons as well~\cite{CHAMPTheory}.

\mysect{Warm D. M.: Non-Minimal Solution}

\subsubsection{Non-Minimal Solution with Warm Dark Matter}

The lifetime of the LSP is clearly an important parameter of the theory, but the mass of the LSP may be even more important. Warm dark matter, a light particle, is also consistent with astronomical data and inflation models~\cite{darkmatter}. An example model that has this feature is Gauge Mediated Supersymmetry Breaking (GMSB), where the process \none $\rightarrow \gamma \widetilde{G}$ is allowed~\cite{gmsbtheory}, where the \grav~is the gravitino, the SUSY partner of the as-yet unobserved mediator of gravity, the graviton. As in the non-minimal cold dark matter solution, the \none~is stable on the time scale of the early universe but decays on the nanosecond timescale to the keV mass \grav, which is still here today, as shown in Figure~\ref{fig: FeynmanDecays}c.

\mysect{Dark Matter and SUSY are Unrelated}

\subsubsection{Other Possibilities, Including the one that Dark Matter and SUSY are Unrelated}

Perhaps the situation is far more complicated than we thought and dark matter and SUSY are unrelated, but SUSY is still a correct description of nature. Maybe there are two (or more) massive, stable, weakly interacting particles in nature~\cite{muriyama}. Possibly Axions play a role~\cite{axions}. In this case, we must look for the most general models including $R$-parity violating terms~\cite{rparity,RefsfromRPV}. These scenarios are also outside the scope of this paper. 

\mysect{Paper Outline}

\subsection{Outline}

With this context we next discuss a number of cosmo-particle searches for sparticles at the Tevatron, as summarized in Table~\ref{tab: Overview}. After a quick experimental introduction in Section~\ref{ExptIssues} we focus on mSUGRA searches for convenience in Section~\ref{mSUGRA}. These include separate direct searches for light squarks and gluinos, gaugino pair production, as well as for Sbottoms and Stops. There are also important indirect searches and direct searches for CHAMPS, which are are particularly sensitive to certain regions of parameter space. In Section~\ref{gmsb}, we discuss GMSB models. 

%In Section~\ref{other}, we look at some other possibilities such as CHAMPS and $R$-Parity violating SUSY searches.

\begin{table}[ht]
\begin{tabular}{| c | c | c | c |} \hline
Model & Search Models & Section & Reference \\ \hline
mSUGRA & $\widetilde{q}$ \& $\widetilde{g}$ in & \ref{multijets} & \protect{\cite{SearchSquarkGluProd}} \\
 & multijets + \met & & \\ \cline{2-4}
 & Gaugino Pairs in & \ref{trileptons} & \protect{\cite{SearchSUSYTrileptSign}} \\
 & trilepton + \met & & \\ \cline{2-4}
 & B$_s \rightarrow \mu ^+ \mu ^-$ & \ref{bsubs} & \protect{\cite{RecentBtoMuMu}} \\ \cline{2-4}
 & Sbottoms & \ref{sbottom} & \protect{\cite{SbottomPairProd, GluMedSbotPro}} \\
 & in $b$-jets + \met & & \\ \cline{2-4}
 & Stops in &  &  \\
 & dilepton & \ref{stop} & \protect{\cite{SearchStopMimick}} \\
 & + jet + \met & & \\ \cline{2-4}
 & CHAMPS: & & \\
 & weakly interacting & \ref{champs} & \protect{\cite{CHAMPSearch}} \\
 & charged, massive, & & \\
 & long-lived particles & & \\ \hline

 Gauge Mediated   & Short-lived \none & \ref{ggmet} & \protect{\cite{GMSBgammagammamet}} \\ 
Symmetry Breaking & in $\gamma\gamma$\met &  & \\ \cline{2-4}
with \none~$\rightarrow \gamma \widetilde{G}$ & Long-lived \none & \ref{delayed} & \protect{\cite{HeavyPartToPhot}} \\ 
  & Delayed Photons & & \\ \hline

% & $R$-Parity Violation &  &  \\
% & in single & \ref{rpv} & \protect{\cite{HighMassRestoLept}} \\ 
% & particle production &  &  \\ \hline
\end{tabular}
\caption{Topics covered in this review.}
\label{tab: Overview}
\end{table}

\section{Experimental Issues}\label{ExptIssues}

Before describing the searches we discus some other experimental issues in looking for the particles of the early universe. The Fermilab Tevatron collides protons and anti-protons with $\sqrt{s}$ = 1.96 TeV and is the high energy frontier until the LHC is turned on. Since a temperature of 100~GeV corresponds to about 10 ps after the Big Bang, the CDF and D\O~detectors allow us to look back at conditions as far back as a picosecond. As this manuscript was created during the Summer of 2009, the Tevatron had delivered about 6 fb$^{-1}$, the detectors had acquired about 5 fb$^{-1}$, and analyzed 2-3 fb$^{-1}$. This paper will cover the results from CDF, and point the reader to the D\O~results for completeness. The Tevatron has resumed taking data as of September 2009, and the plan is to run for at least a year. There are rumors of having it run until 2012, as it keeps delivering more and better data as time goes on. This, of course, critically depends on the LHC's progress; for more on LHC start-up issue, see Appendix A. 

A detailed description of the CDF detector can be found elsewhere~\cite{DetectorStuff}. It is the prototypical collider detector with concentric sub-detectors used for particle identification and 4-momentum measurement. From the inside out it contains a silicon tracker, a central tracker, electromagnetic calorimeters, hadronic calorimeters, and muon chambers.  After more than 20 years of experience there now exists well understood and standardized identification methods for electrons, muons, taus, photons, jets, $b$-jets, and missing energy. The missing energy is particularly important for the identification of dark matter candidates. As a weakly interacting neutral particle, it would, if produced, leave the detector without interacting, like the SM neutrino, and create the missing energy imbalance we call \met. Other new detector techniques have come online for Run~II, including 25 ps charged particle timing~\cite{TOFRef}, 500 ps photon timing~\cite{emtim}, and powerful new \met~resolution models~\cite{ggxprd}.

\mysect{Aside Before We Begin...}

\newcommand{\MyAside} {It is important to note that most of the analyses presented in this paper will look like they were easy in hindsight. At the time of this writing it's 2009 and Run~II is eight years into into it's lifetime. These analyses are a lot more difficult than they look and take a lot longer than they, perhaps, should. Run~II at the Tevatron was effectively started in 2001, the first indirect SUSY search paper appeared in print in 2004~\cite{firstIIbs} with the first direct search paper following suit in 2005~\cite{FirstIIggMet}. We also note that the first direct search, in $\gamma \gamma +$\met, was the ``simplest'' search analysis with the photons dominating the calorimeter response, minimizing the \met~resolution and the reconstruction pathologies, crucial for robust SUSY searches. Although not strictly true, both analyses were also largely repeats of Run~I analyses~\cite{RunIbs, eeggmet}. How quickly will the LHC and its detectors come up? CDF was a known detector and the Tevatron was a known accelerator, and there were five times fewer collaborating scientists. While there's only a sample of one and there will be great pressure to get results out quickly, a historical perspective says that since the CDF detector was known better the LHC results will come out more slowly relative to first collisions. The same time-ordering is to be expected since the Tevatron/CDF combination had working/known accelerator and well tested Monte Carlo simulations of the collision process and detector response. As for having five times as many people working on CMS/ATLAS than on CDF/D\O, only time will tell if this makes things go quicker or more slowly, but compelling arguments can be made either way.}

\section{Minimal Supergravity Models: mSUGRA}\label{mSUGRA}

\mysect{Overview}

Models of SUSY where the symmetry breaking is mediated by the gravity sector are known as Supergravity models. Most have some simplifying assumptions and the most commonly used minimal model is known as minimal Supergravity or mSUGRA~\cite{mSUGRA}. In this model the scalars are assumed to have a common mass, $m_0$, and sfermions have a common mass, $m_{1/2}$, at the unification scale. To first order most sparticle masses scale with these parameters. At $m_{GUT}$ there are three other free parameters that determine the sparticle masses at the EWK scale. They are tan$\beta$, which is the ratio of the Higgs Vacuum Expectation Values (VEV), which will be important later, $A_0$, the common trilinear coupling factor, and sgn$(\mu)$, where $\mu$ is the Higgs mass parameter. While this model may be too simple to be true, it provides a useful phenomenological benchmark for use in both focusing searches and interpreting the sensitivity of the searches in the case of null results.

\mysect{The Sparticle Masses}

In a typical mSUGRA scenario available at Tevatron energies, and taking into account experimental constraints, the squarks and gluinos are the heaviest sparticles, the first and second generation sparticles are mass degenerate, and the lightest neutralino is the LSP, making it a good cold dark matter candidate. The typical mass relations are approximately $m_{\widetilde{\chi}_1^0}$: $m_{\widetilde{\chi}_2^0}$ : $m_{\widetilde{\chi}_1^{\pm}}$ : $m_{\widetilde{g}}$ = 1 : 2 : 2 : 6 and $m_{\widetilde{q}} \cong m_{\widetilde{g}}$. The masses of the stop, sbottom, and stau are very dependent on tan$\beta$ and for large values, tan$\beta \geq 10$, can be very light. This case has the aesthetically pleasing aspect that, if true, as in the SM, members of the third generation would have very different masses than members of the other generations.

\mysect{High Tan\bg~vs. Low Tan\bg}

\begin{figure}[b]
\figurehelper{\includegraphics[bb=104pt 65pt 373pt 273pt, width=8.66cm]{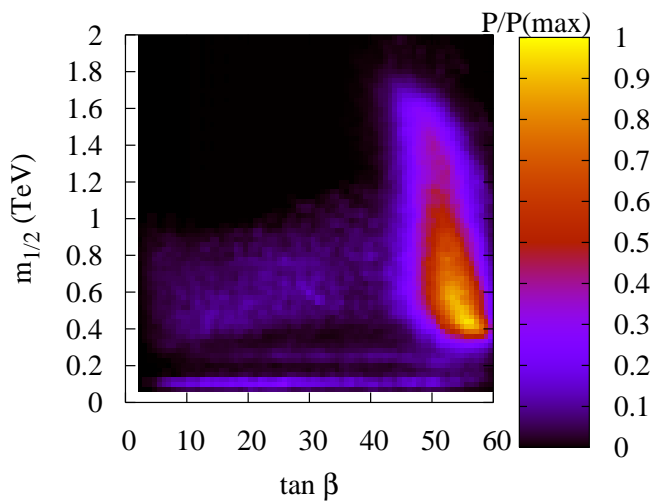}}
\caption{A likelihood fit of mSUGRA in the $m_{1/2}$ vs tan$\beta$ plane to the experimental data. The data clearly favor high tan$\beta$ in these types of models. Taken from Ref.~\protect{\cite{HighTanB}}}
\label{fig: MassFitsHighTanB}
\end{figure}

While it's important not to place too much emphasis on a particular SUSY model, likelihood fits of mSUGRA to existing data including the current Higgs mass limits, $g-2$ and other experimental data in the plane of $m_{1/2}$ and tan$\beta$ show better fits with high tan$\beta$~\cite{HighTanB}, as shown in Figure~\ref{fig: MassFitsHighTanB}. This strongly suggests that to have better coverage for a variety of models that we do a range of searches to cover both low tan$\beta$ and high tan$\beta$. 

\mysect{Golden Search Channels}

Given these inputs our benchmark cold dark matter searches at the Tevatron can be broken down into a number of types which have distinct advantages and disadvantages, depending on what is realized in nature. They are:

\begin{itemize}
\item {\it Direct production of squarks and gluinos}: Like quarks and gluons, as strongly interacting particles, squarks and gluinos should have the largest production cross sections per unit mass at the Tevatron. However, as the production cross sections also drop rapidly as a function of sparticle mass, they may be too heavy to be produced at the Tevatron. An additional complication is that since they typically produce multijet + \met~final states they are difficult to separate from the large QCD backgrounds with fake \met.
\item {\it Direct production of gauginos}: The gauginos are typically lighter than the colored objects which could make their production cross section large. However, they have EWK production mechanisms, which make the cross sections smaller per GeV/$c^2$ of mass. If the lightest neutralino is about, or above, 75~GeV/$c^2$  (from LEP~\cite{LimitsLEP}) then, from the above mass relations, the squarks and gluinos would be too heavy to be produced at the Tevatron. However, gaugino pairs could be produced at observable levels. A second advantage of gaugino pair production is that they often have leptonic final states, for which there are smaller backgrounds.

\item {\it Third generation searches}: At high tan$\beta$ the stop and sbottom masses might be light enough to make them easier to produce. Often they produce final state $b$ or $t$ quarks which could be identified. Similarly, gaugino branching fractions to $\tau$'s can rise to 100\% as the stau gets light making their signature also very distinct. On the flip-side, the purity and efficiency for $\tau$ and $b$-quark identification are not at the same level as their first and second generation counterparts.

\item {\it Indirect searches via sparticles in loops}: These searches measure the branching ratios of SM particle decays for evidence of sparticles in loops. If all the sparticles are too heavy to be directly produced, this may be our best, or only, bet.

\item {\it CHAMPS searches}: If the next-to-lightest SUSY particle (NLSP) is long-lived, then, as in indirect searches, there will be no direct \met~signature. In this case, this again may be our best, or only, bet. 

\end{itemize}

We consider all these types separately to maximize coverage and start with low tan$\beta$, moving to searches with higher tan$\beta$. We again note that, as for many CDF results, there are typically comparable D\O~search results. We will not cover these individually, but reference them for completeness.

\subsection{Low Tan\bg~Searches}

\subsubsection{Squarks \& Gluinos}\label{multijets}

\mysect{Squark/Gluino Searches: Multijet + Met}

\begin{figure}[b]
\figurehelper{\includegraphics[width=8.66cm, height=6cm]{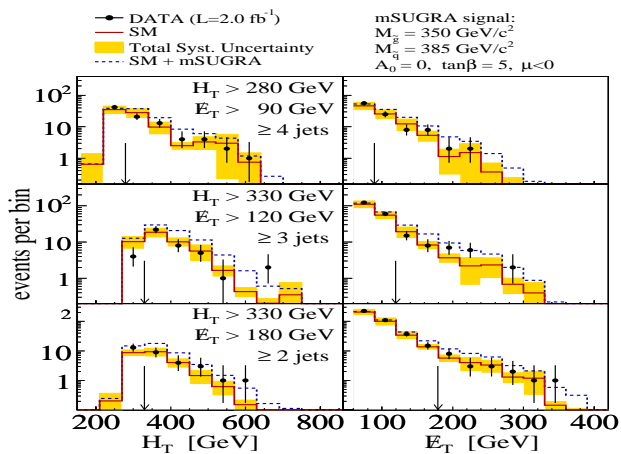}}
\caption{The $H_T$ and \met~distributions in (Top) 2 jet + \met, (Middle) 3 jet + \met, and (Bottom) 4 jet + \met~events along with the background expectations and an example distribution for squark and gluino production. Taken from~\protect{\cite{SearchSquarkGluProd}}.}
\label{fig: EventsVsMet}
\end{figure}

For squark and gluino masses available at the Tevatron, there are three main production diagrams: $\widetilde{q}\widetilde{q}$, $\widetilde{g}\widetilde{g}$, and $\widetilde{q}\widetilde{g}$. As the two masses are expected to be similar, small differences between them can produce final states which can vary significantly. Specifically, the decays are expected to be  $\widetilde{q} \rightarrow q\widetilde{\chi}_1^0$ and $\widetilde{g} \rightarrow q\widetilde{q} \rightarrow q(q\widetilde{\chi}^0_1)$ where the intermediate squark is real or virtual, depending on the masses. These produce three separate final states: two jets (for $m_{\widetilde{g}} \gg m_{\widetilde{q}}$), three jets (for $m_{\widetilde{g}} \cong m_{\widetilde{q}}$), and four jets (for $m_{\widetilde{g}} \ll m_{\widetilde{q}}$), where the weakly interacting \none's leave the detector creating \met. This suggests a multijet + \met~analysis. CDF now uses a new, optimized strategy with 2.0~fb$^{-1}$ of data and searches for two jets + \met, three jets + \met, and four jets + \met~events separately, but with a unified analysis to give the best coverage~\cite{SearchSquarkGluProd}. The analysis selects multijet events with large \met~and $H_T$, (where $H_T$ is defined as the scalar sum of the jets and \met) as well as restrictive kinematics to ensure well measured \met~in events. Despite huge QCD production cross sections, the sophistication of the analysis techniques have reduced the QCD backgrounds with fake \met~to the 25\% level. The rest is $t\bar{t}$ and other EWK processes with final state neutrinos which produce real jets and \met. Figure~\ref{fig: EventsVsMet} shows the $H_T$ and \met~distributions in large $H_T$ events for two, three, and four jet events. Table~\ref{tab: MetPetJets} gives the results of the counting experiments.

\begin{table}[hbt]
\begin{tabular}{| c | c | c | c |} \hline
 & \multicolumn{3}{| c |}{Jets} \\ \cline{2-4}
 & 2 & 3 & 4 \\ \hline
Selections & H$_T$ $>$ 330~GeV & H$_T$ $>$ 330~GeV & H$_T$ $>$ 280~GeV \\
 & \met~$>$ 180~GeV & \met~$>$ 120~GeV & \met~$>$ 90~GeV \\ \hline
Expected & 16 $\pm$ 5 & 37 $\pm$ 12 & 48 $\pm$ 17 \\
SM Events& & & \\ \hline
Observed & 18 & 38 & 45 \\ 
Events &  &  & \\ \hline
\end{tabular}
\caption{The number of events in the multijet + \met~search for squarks and gluinos. For example signals see Figure~\protect\ref{fig: EventsVsMet}. Taken from~\protect{\cite{SearchSquarkGluProd}}.}
\label{tab: MetPetJets}
\end{table}

\mysect{Limits}

\begin{figure}[bth] 
\figurehelper{\includegraphics[width=9.5cm]{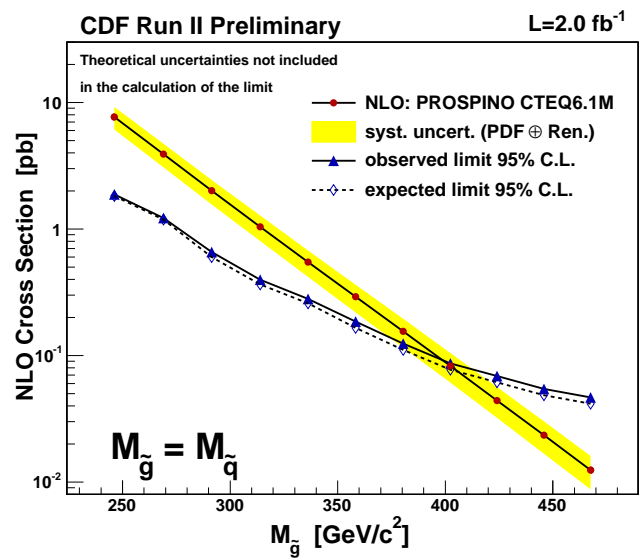}}
\caption{The 95\% confidence level cross section upper limits for the case $m_{\widetilde{g}} = m_{\widetilde{q}}$ from the unified multijet + \met~search. Taken from~\protect{\cite{SearchSquarkGluProd}}.}
\label{fig: CrossSectionVsMasses}
\end{figure}

\begin{figure} [ht]
\figurehelper{\includegraphics[width=8.66cm]{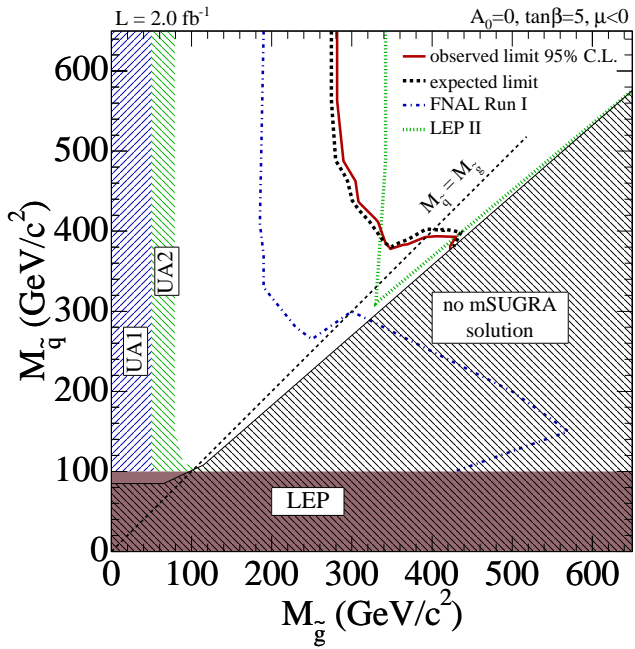}}
\caption{The 95\% confidence level exclusion region in the squark vs. gluino mass plane from the multijet + \met~search. Taken from~\protect{\cite{SearchSquarkGluProd}}.}
\label{fig: SquarkGluinoMassLimits}
\end{figure}

\begin{figure}
\figurehelper{\includegraphics[width=9.2cm]{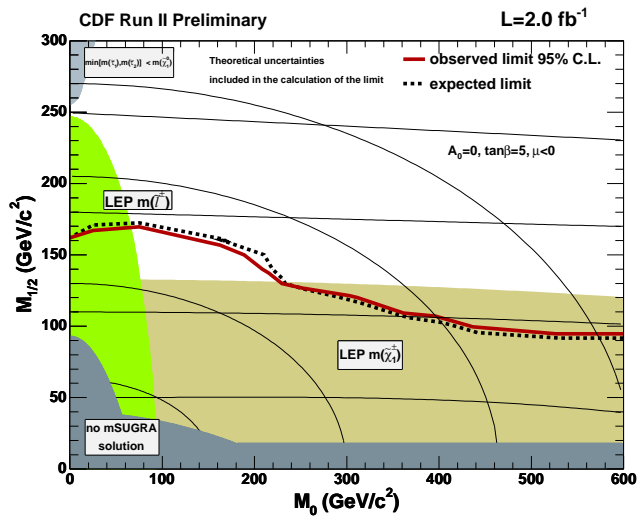}}
\caption{The 95\% confidence level exclusion region in the $m_0$ vs. $m_{1/2}$ mSUGRA mass parameter plane from the multijet + \met~search for squarks and gluinos. Taken from~\protect{\cite{SearchSquarkGluProd}}.}
\label{fig: LimitmSUGRAParameters}
\end{figure}

Since there is no evidence for new physics, 95\% confidence level limits are set. Setting limits can be done in a number of different ways to illustrate the experimental sensitivity to new physics as more and more model dependence is added into the results. Cross section limits have a significant advantage in that they are fairly model independent and essentially reflect a balance between an acceptance model that comes from the kinematics of the particles produced and their decays, and the backgrounds and the ability to reject them. The observed and expected cross section upper limits for the squark and gluino searches are shown in  Figure~\ref{fig: CrossSectionVsMasses} for the special case of $m_{\widetilde{g}} \cong m_{\widetilde{q}}$. If more model dependence is added by including the production cross section (i.e., more physics in the interaction vertices) mass limits can be extracted from where the cross section limits are below the production cross section. This is essentially where the two lines in Figure~\ref{fig: CrossSectionVsMasses} cross, but also takes into account systematic uncertainties on the production cross section. The exclusion region in the squark vs. gluino mass plane is  shown in Figure~\ref{fig: SquarkGluinoMassLimits}. Some notes are in order: $m_{\widetilde{g}} < 280$~GeV/$c^2$ is always excluded and $m > 392$~GeV/$c^2$ is excluded when $m_{\widetilde{g}} = m_{\widetilde{q}}$. A typical last step in the process is to invoke the mechanism of SUSY breaking to realize the sensitivity to model parameters. For mSUGRA, for the parameter choice used, the exclusion region is shown in Figure~\ref{fig: LimitmSUGRAParameters} in the $m_0$ vs. $m_{1/2}$ plane. Of particular interest is that the limits go beyond those from LEP~\cite{LimitsLEP} for mSUGRA models in the region $75 < m_0 < 250$~GeV/$c^2$ and $130 < m_{1/2} < 170$~GeV/$c^2$. Comparable results from D\O~can be found in Ref.~\cite{D0SquarkGlu}.

\subsubsection{Gaugino Pair Production}\label{trileptons}

\mysect{Gaugino Production: Multilepton + Met}

Direct gaugino production at the Tevatron has EWK production mechanisms, but can dominate the sparticle production cross section if the squarks and gluinos are too heavy to be produced at the Tevatron. It is favored in mSUGRA scenarios where the Higgs mass limits and lightest neutralino mass limits from LEP are taken into account~\cite{LEPHiggslimits}. The dominant production diagrams are typically \conep \conem~and \conep \ntwo~production, which are analogous to SM $WW$ and $WZ$ production. A common leptonic decay mode of the chargino is typically \conep $\rightarrow W^*\widetilde{\chi}^0_1 \rightarrow (l\nu)$\none, which means that chargino pairs will produce a two lepton + \met~final state, which is difficult to distinguish from SM backgrounds. The dominant decays of the neutralino are $\widetilde{\chi}_2^0 \rightarrow \widetilde{l}l \rightarrow \left(l \widetilde{\chi}_1^0 \right) l$ and $\widetilde{\chi}_2^0 \rightarrow Z^* \widetilde{\chi}_1^0 \rightarrow (l^+ l^- ) \widetilde{\chi}_1^0$, depending on the $m_{\widetilde{\chi}^0_2}-m_{\widetilde{l}}$ mass differences. Thus, \conep \ntwo~production can result in a final state of three leptons + \met, which has very few SM backgrounds.

\mysect{Unified Gaugino Production Analysis}

\begin{figure}
\figurehelper{\includegraphics[width=8.66cm]{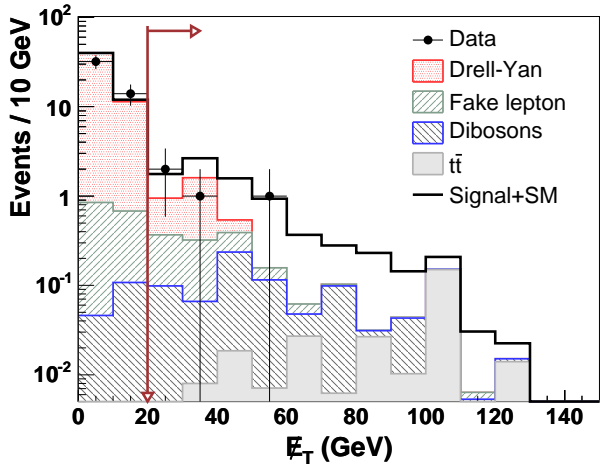}}
\caption{ The \met~distribution in 2 Tight + 1 Track events as part of the search for gaugino pair production in the trilepton + \met~final state. Taken from~\protect{\cite{SearchSUSYTrileptSign}}.}
\label{fig: GauginoPairLimits}
\end{figure}

Most previous Tevatron trilepton + \met~searches at low tan$\beta$ considered only the $eee$, $ee\mu$, $e\mu \mu$, or $\mu \mu \mu$ final states. Rather than focus on $e$'s and $\mu$'s separately, the latest search, with 2.0 fb$^{-1}$ of data, considers five separate event quality types in a unified analysis to expand the coverage and include final state hadronic $\tau$'s~\cite{SearchSUSYTrileptSign}. ``Tight'' lepton identification helps select electrons or muons in a manner that produces very pure samples (i.e. low backgrounds from jets that fake leptons) at the cost of efficiency. ``Loose'' lepton identification accepts more electrons and muons with better efficiency at the cost of purity, i.e. has more background from fake leptons. A third lepton category is to only require an isolated, charged particle track to identify  an electron, muon or single-prong hadronic tau decay. Note that leptons considered tight cannot simultaneously be identified as loose or a track. Similarly, a lepton identified as loose-but-not-tight cannot be identified as a track. The five non-overlapping categories, along with the results of the  background expectations and the experimental results are shown in Table~\ref{tab: UniGauginoPair}. Figure~\ref{fig: GauginoPairLimits} shows the \met~distribution of the 2 Tight + 1 track final state. There is no evidence for new physics.

\mysect{mSUGRA Exclusion Region}

\begin{figure}
\figurehelper{\includegraphics[width=8.66cm]{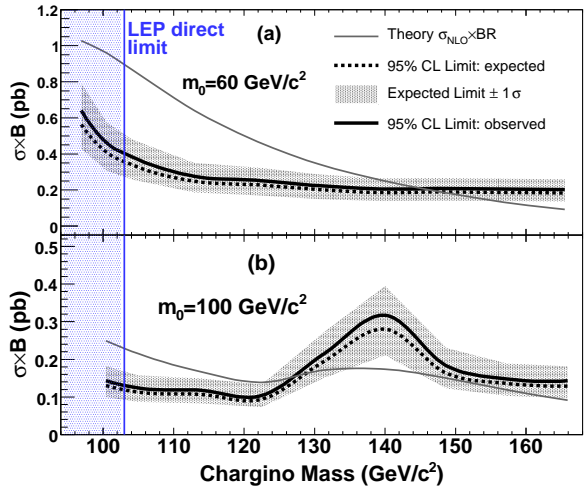}}
\caption{Cross section limits on chargino-neutralino production and decay in the trilepton + \met~search for two different values of $m_0$. Note that in the bottom plot the limits get less restrictive near the $m_{\widetilde{\tau}} = m_{\widetilde{\chi}_2^0}$ mass point. Taken from~\protect{\cite{SearchSUSYTrileptSign}}.}
\label{fig: CharginoMassLimits}
\end{figure}

\begin{figure} [ht]
\figurehelper{\includegraphics[width=8.66cm]{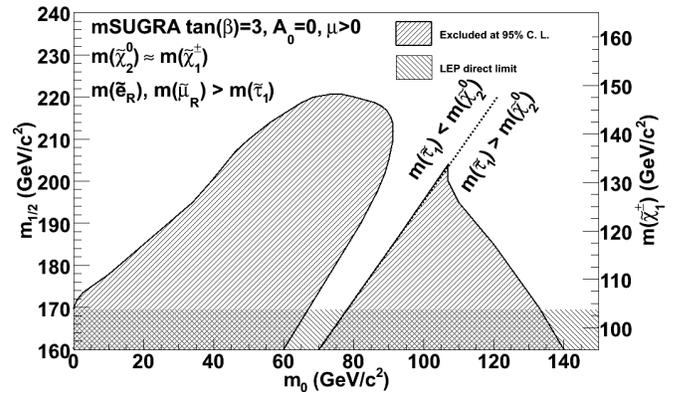}}
\caption{The mSUGRA exclusion region from the search for chargino-neutralino production and decay in the trilepton + \met~final state~\cite{SearchSUSYTrileptSign}. Note the unexcluded region in the middle which occurs when the stau mass is approximately equal to the \ntwo~mass, producing a third lepton which has a $P_T$ which is so low as to be undetectable. For more detail, see~\protect{\cite{SunilsPaper}}.}
\label{fig: TrileptonMassFigure}
\end{figure}

The interpretation of the data is not only gaugino mass dependent, but is also determined by the \ntwo~decay branching ratios which are very slepton mass dependent. Figure~\ref{fig: CharginoMassLimits} shows the 95\% confidence level upper limits on the cross section as a function of the chargino mass in the large and small slepton mass scenarios. Taking these effects into account we find the mSUGRA exclusion region shown in Figure~\ref{fig: TrileptonMassFigure}. The middle region, where there is no sensitivity, is where the stau mass changes from being above the \ntwo~mass to below. In this region there is such a small mass difference between the \ntwo~and slepton masses that there is very little energy for one of the lepton P$_T$ in \ntwo~$\rightarrow \widetilde{l}l$ decays, making it undetectable in the detector. Reference~\cite{SunilsPaper} gives a method for model-independent interpretation along with numerical formulae for generalizing the Tevatron trilepton search results. Comparable D\O~search results can be found in~\cite{D0Trilepton}.

\begin{table}[hbt]
\begin{tabular}{| l | c | c |} \hline
Channel & Background & Obs \\ \hline
3 Tight & 0.49 $\pm$ 0.04 $\pm$ 0.08 & 1 \\ \hline
2 Tight + 1 Loose & 0.25 $\pm$ 0.03 $\pm$ 0.03 & 0 \\ \hline
1 Tight + 2 Loose & 0.14 $\pm$ 0.02 $\pm$ 0.02 & 0 \\ \hline
2 Tight + 1 Track & 3.22 $\pm$ 0.48 $\pm$ 0.53 & 4 \\ \hline
1 Tight + 1 Loose + 1 Track & 2.28 $\pm$ 0.47 $\pm$ 0.42 & 2 \\ \hline
\end{tabular}
\caption{The five final state configurations in the united search for chargino-neutralino production and decay in the trilepton + \met~final state. Taken from~\protect{\cite{SearchSUSYTrileptSign}}.}
\label{tab: UniGauginoPair}
\end{table}

\subsection{High Tan\bg~Searches}

\subsubsection{Indirect Searches}\label{bsubs}

\mysect{Indirect Search: B$_s \rightarrow \mu \mu$}

\begin{figure} 
\figurehelper{\includegraphics[width=8.66cm]{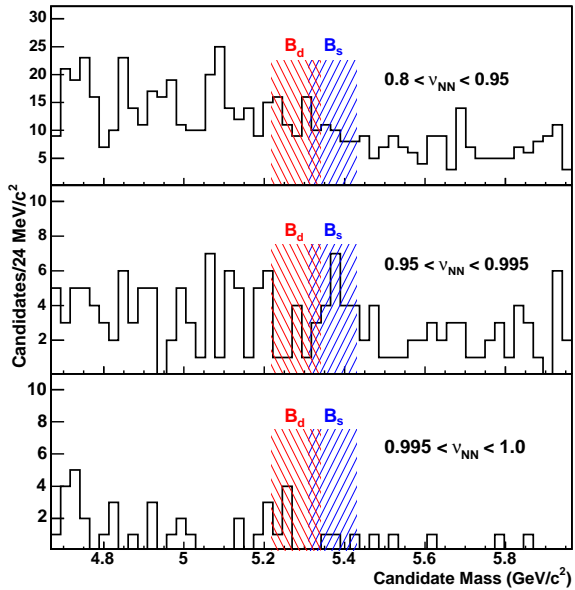}}
\caption{The $\mu \mu$ invariant mass with various neural net requirements from the search for $B_s \rightarrow \mu^+ \mu^-$. The neural net bins are define such that values close to 1.0 are $B_s \rightarrow \mu^+ \mu^-$-like and values close to 0.0 are background-like. Taken from~\protect{\cite{RecentBtoMuMu}}.}
\label{fig: CandVsCandMass}
\end{figure}

The search for $B_s \rightarrow \mu ^+ \mu ^-$ is perhaps the most sensitive to SUSY at high tan$\beta$ since sparticles can show up in loops as the branching fraction rises as tan$^6 \beta$~\cite{DetBs}. In the SM, the decay of $B_s \rightarrow \mu^+ \mu^-$ is heavily suppressed and has a prediction of  $BR_{SM} (B_s \rightarrow \mu^+ \mu^-) = (3.5 \pm 0.9) \times 10^{-9}$~\cite{SMbsubs}. This was the first SUSY search out of CDF in Run~II~\cite{firstIIbs} and since then has been heavily optimized using neural net techniques and 2.0~fb$^{-1}$ of data~\cite{RecentBtoMuMu}. The backgrounds are combinatorial and estimated from data using sideband techniques. Of particular note is that the backgrounds have not successfully been estimated from Monte Carlo simulations, which makes predictions for sensitivity at the LHC precarious. An $a$ $priori$ analysis considers three neural net bins, shown in Figure~\ref{fig: CandVsCandMass}, and 5 separate mass bins; each of which is compared to expectations. The neural net bins are defined such that values close to 1.0 are $B_s \rightarrow \mu^+ \mu^-$-like and values close to 0.0 are background-like. 

\begin{figure} 
\figurehelper{\includegraphics[bb=12pt 61pt 545pt 653pt, width=8.66cm]{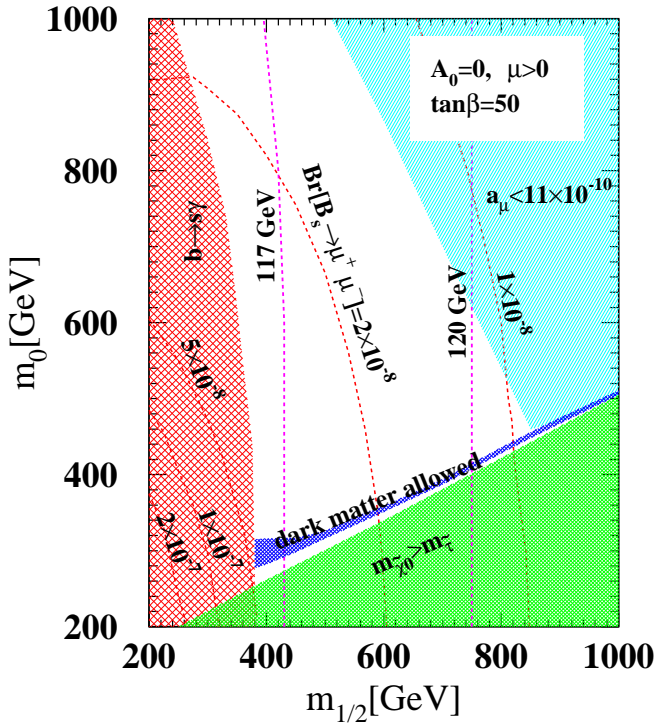}}
\caption{The mSUGRA region for large tan$\beta$ and the exclusions from various other experiments. Taken from~\protect{\cite{DetBs}}. The current limits from Ref.~\cite{RecentBtoMuMu} are $5.8 \times 10^{-8}$.}
\label{fig: BranchingRatioLimits}
\end{figure}

Since there is no evidence for new physics a 95\% confidence level upper limit of $5.8 \times 10^{-8}$ is set on the branching fraction. Comparable results from D\O~can be found in~\cite{D0BsToMuMu}. In mSUGRA, cosmology constraints point to high tan$\beta$ and the co-annihilation region, where $\widetilde{\tau}$-\none~mass difference is expected to be less than 20~GeV/$c^2$/$c^2$~\cite{DetBs}, as shown in Figure~\ref{fig: BranchingRatioLimits}. These limits correspond to an exclusion of $m_{1/2}$ below 380~GeV/$c^2$ and gluino masses below 925~GeV/$c^2$. As expected these are well above the direct search mass limits and are the world's most sensitive. 

\subsubsection{Sbottom Searches}\label{sbottom}

\mysect{Sbottom Searches}

For high tan$\beta$ the bottom squarks can be lighter than the other squarks and may well be the best direct search method. Also, powerful $b$-tagging techniques, developed for top quark properties measurements~\cite{btagging}, enable complex searches for events from $\widetilde{b} \rightarrow b \widetilde{\chi}^0_1$ decays producing $b$-jets + \met~final state. There are two main types of sbottom searches.  The first is from gluino pair production via $\widetilde{g}\widetilde{g} \rightarrow (\widetilde{b}b)(\widetilde{b}b) \rightarrow (b\bar{b} \widetilde{\chi}_1^0)(b\bar{b} \widetilde{\chi}_1^0)$, the second is direct sbottom pair production $\widetilde{b} \widetilde{\bar{b}} \rightarrow (b \chi _1^0)(b \chi _1^0)$. Direct sbottom searches in Run~II were published in 2007 by CDF~\cite{ScalBottGluDecays}, with an update in 2009~\cite{MonicaSbottom} and in 2006 by D\O~\cite{D0SbottomDirect}. 

\mysect{After all 3 Sparticle Production}

\begin{figure} 
\figurehelper{\includegraphics[width=8.66cm]{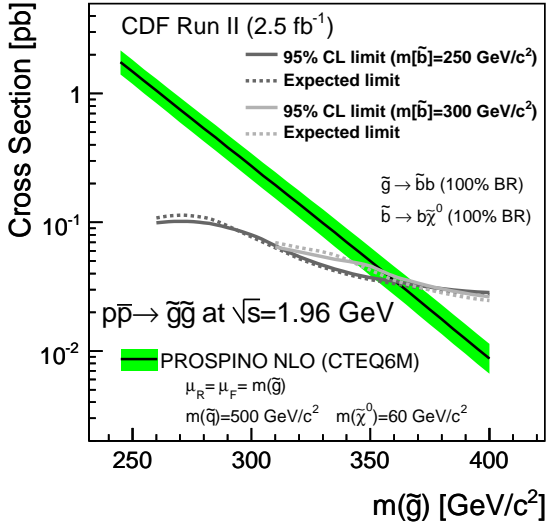}}
\caption{The 95\% confidence level cross section limits for pair production of gluinos that decay via $\widetilde{g} \rightarrow b\widetilde{b} \rightarrow b(b\widetilde{\chi}^0_1)$ into the $b$-jets +\met~final states. Taken from~\protect{\cite{GluMedSbotPro}}.}
\label{fig: CrossSectionVsGluinoMass}
\end{figure}

\begin{figure}[htb]
\figurehelper{\includegraphics[width=8.66cm]{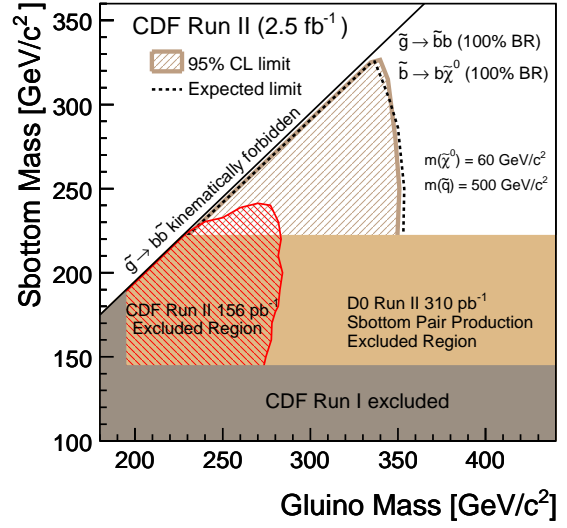}}
\caption{The exclusion regions in the sbottom mass vs. gluino plane from searches in the $b$-jets + \met~final state for gluino mediated sbottom production. Also shown are results from searches for direct sbottom production. Taken from~\protect{\cite{GluMedSbotPro}}.}
\label{fig: SbottomMassVsGluinoMass}
\end{figure}

New results have recently appeared in gluino-mediated scenario searches in the multi-$b$-tag + \met~final state~\cite{SbottomPairProd} with 2.5~fb$^{-1}$ of data. The gluino production searches are complementary to the direct sbottom searches which are gluino mass independent, but are only sensitive to lower mass sbottoms. Two separate optimizations are performed to cover the cases when the gluino-sbottom mass difference is large, producing larger $E_T$ $b$-jets in the final state, and when the mass differences are smaller and the $b$-jets are less energetic. After all requirements, including $b$-tagging, large $H_T$, and large \met~and neural net techniques, the backgrounds are roughly half QCD and half top quark and EWK production. The results of the two counting experiments are given in Table~\ref{tab: SbottomData}. Since there is no evidence for new physics, cross section limits on the gluino are shown in Figure~\ref{fig: CrossSectionVsGluinoMass}. Figure~\ref{fig: SbottomMassVsGluinoMass} shows the exclusions in the sbottom mass vs. gluino mass plane along with results from Ref.~\cite{ScalBottGluDecays}. The most recent results in this final state from D\O~can be found in Ref.~\cite{D0SbottomRecent}.

\begin{table}[htb]
\begin{tabular}{| c | c | c |} \hline
Background & Large $\delta m$ & Small $\delta m$ \\
Source & Optimization & Optimization \\ \hline
Electroweak boson & 0.17 $\pm$ 0.05 & 0.5 $\pm$ 0.3 \\
backgrounds &  &  \\ \hline
Top-quark & 1.9 $\pm$ 1.0 & 0.6 $\pm$ 0.4 \\ \hline
Light-flavor jets & 1.0 $\pm$ 0.3 & 0.6 $\pm$ 0.1 \\ \hline
Heavy flavor  & 1.6 $\pm$ 0.8 & 0.7 $\pm$ 0.3 \\ 
multijets &  &  \\ \hline
Total expected SM & 4.7 $\pm$ 1.5 & 2.4 $\pm$ 0.8 \\ \hline
Observed & 5 & 2 \\ \hline
\end{tabular}
\caption{The results of the counting experiments in the $b$-jets + \met~search for gluino-mediated production of sbottoms pairs. Large $\delta m$ refers to scenarios where $m_{\widetilde{g}} \cong$ 335~GeV/$c^2$ and $m_{\widetilde{b}} \cong$ 260~GeV/$c^2$, which produces high $E_T$ final state $b$-jets.
Small $\delta m$ refers to scenarios where $m_{\widetilde{g}} \cong$ 315~GeV/$c^2$ and $m_{\widetilde{b}} \cong$ 335~GeV/$c^2$, which will produce low $E_T$ final state jets. Taken from~\protect{\cite{GluMedSbotPro}}.}
\label{tab: SbottomData}
\end{table}

\subsubsection{Stop Searches}\label{stop}

\mysect{Stop Searches}

The lightest stop mass eigenstate, $\widetilde{t_1}$, can be significantly lighter than all the other non-LSP sparticles due to a large top-Yukawa coupling, especially at large tan$\beta$. There have been many searches for pair production and decay of $\widetilde{t_1}$ in different decay modes including via $\widetilde{t_1} \rightarrow \widetilde{\chi}_1^+ b \rightarrow (l\nu \widetilde{\chi}_1^0)b$, $\widetilde{t_1} \rightarrow \widetilde{\chi}_1^+ b \rightarrow (jj \widetilde{\chi}_1^0)b$, and $\widetilde{t} \rightarrow c$\none. 
Searches where both stops decaying $\widetilde{t} \rightarrow c$\none~ were published in~\cite{ScalBottGluDecays}, and updated in 2009~\cite{MiguelStop}. A similar search from D\O~can be found in Ref.~\cite{D0StopCharm}. A search via the chargino channels where one stop decays leptonically and the other hadronically was carried out by D\O~in 2006~\cite{D0StopAdmix}.
Recently a new search in the dilepton + jets + \met~final state signature was performed. Special motivation for this channel  comes from the Run~I observation that some of the dilepton events didn't ``look'' like top quark pair production and decay~\cite{Hall}. Similarly, an admixture of stop events among the top quark events would effect the kinematics of the final state objects and would explain differences in the top mass measurements in the jets+\met, the lepton+jets+\met, and the dilepton+jets+\met~final states as seen in early Run~II~\cite{TopMassMeasurements}. 

\begin{figure} 
\figurehelper{\includegraphics[width=8.66cm]{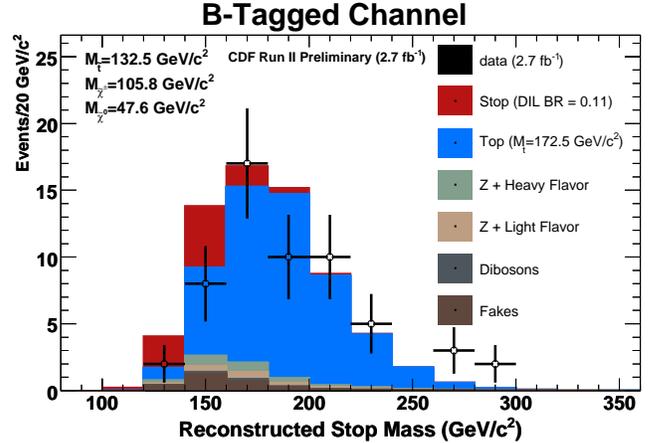}}
\caption{The mass distribution of dilepton + \met~+ jet events reconstructed under the stop hypothesis. Taken from~\protect{\cite{SearchStopMimick}}.}
\label{fig: EventsVsStopMass}
\end{figure}

\begin{figure} 
\figurehelper{\includegraphics[width=8.66cm, height=6cm]{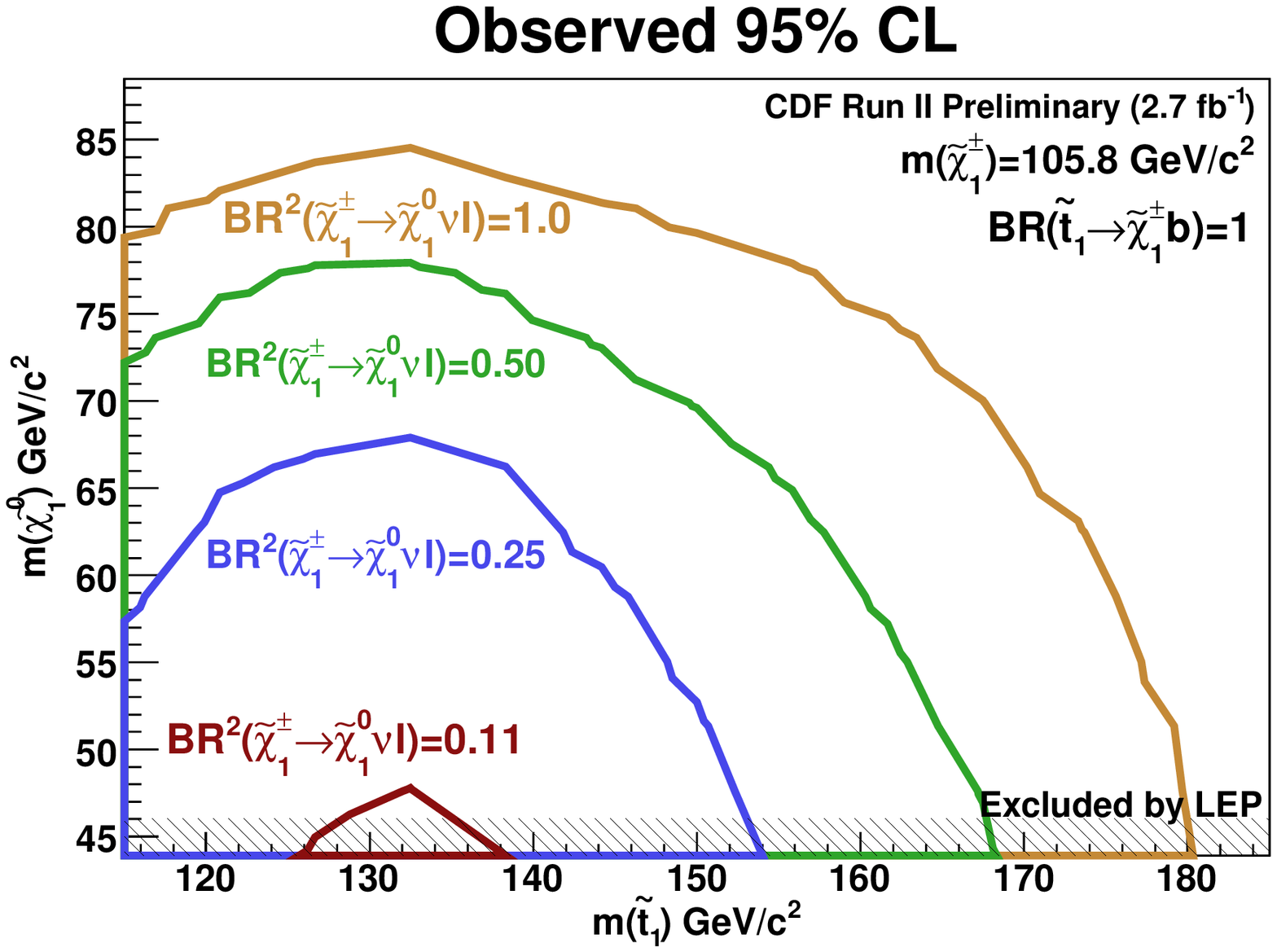}}
\figurehelper{\includegraphics[width=8.66cm, height=6cm]{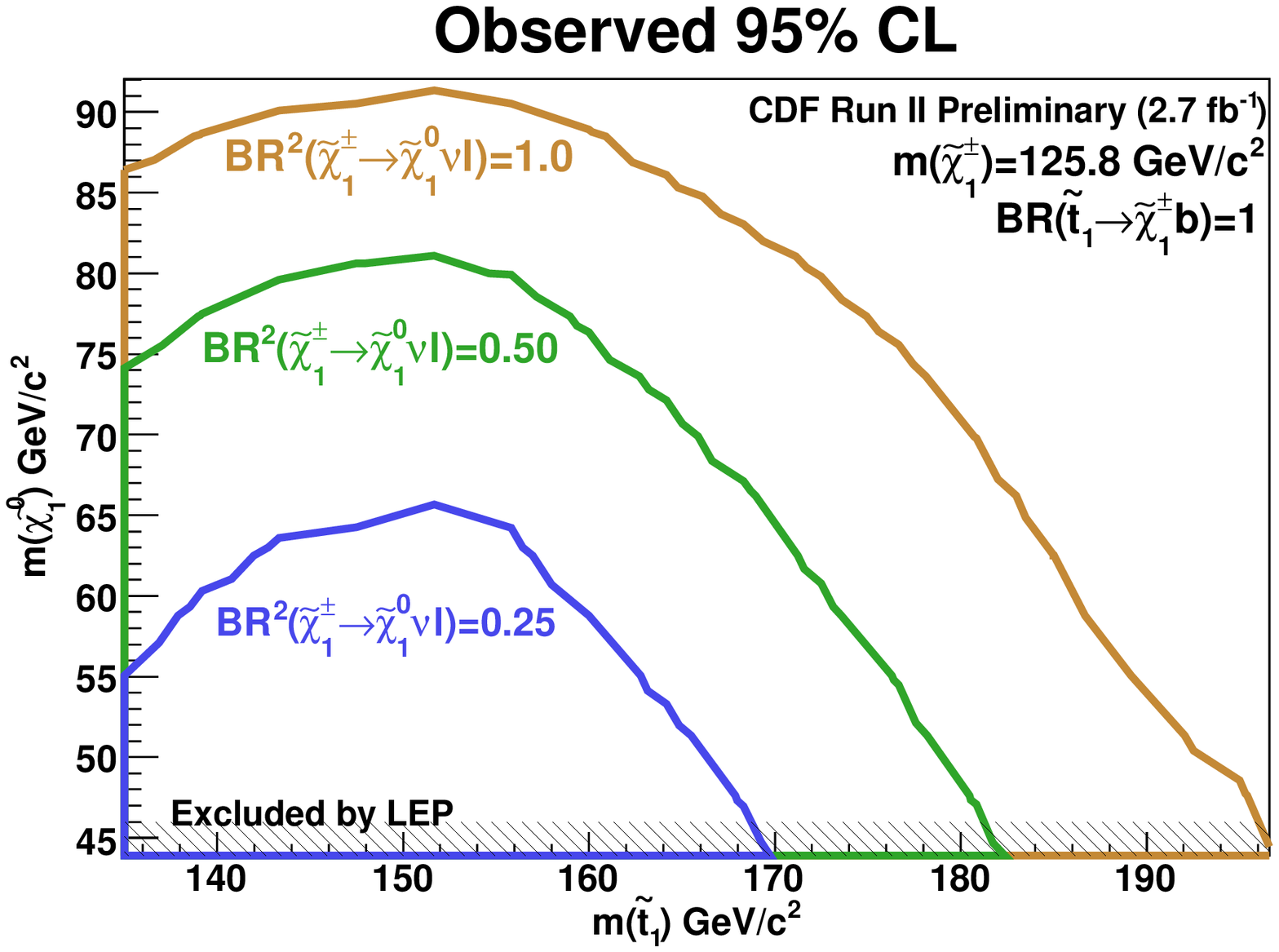}}
\caption{The 95\% confidence limits on stop production and decay as a function of chargino and $\widetilde{t_1}$ masses from the search for stops in the dilepton + jets +\met~final state. Taken from~\protect{\cite{SearchStopMimick}}.}
\label{fig: CharginoMassVsStopMass}
\end{figure}

A powerful new method~\cite{SearchStopMimick} does a fit of the kinematics of all dilepton + dijet + \met~events 
after minimal top quark event rejection requirements to search for stop events mixed in with the top quark events. In this search in 2.7~fb$^{-1}$ of data the $\widetilde{t_1}$, \chargino, and \none~masses are allowed to vary in the fit. Figure~\ref{fig: EventsVsStopMass} shows the best fit $\widetilde{t_1}$ mass distribution along with the background predictions. As there is no evidence of new physics, branching ratio limits are set as a function of the $\widetilde{t_1}$ and $\widetilde{\chi}^+_1$ masses. The results are shown in Figure~\ref{fig: CharginoMassVsStopMass} and are the first exclusions on these sparticles in the  
$\widetilde{t_1} \rightarrow \widetilde{\chi}_1^+ b \rightarrow (l\nu \widetilde{\chi}_1^0)b$ channel.
%masses of $\widetilde{t_1}$ and $\widetilde{\chi}_1^0$ for several values of $m_{\widetilde{\chi}_1^+}$ and branching ratio of $\widetilde{\chi}_1^+ \rightarrow l\nu \widetilde{\chi}_1^0)b$. 

\subsection{CHAMPS}\label{champs}

\mysect{Stable Charged Sparticles (CHAMPS)}

\begin{figure}[ht]
\figurehelper{\includegraphics[width=8.66cm, height=6cm]{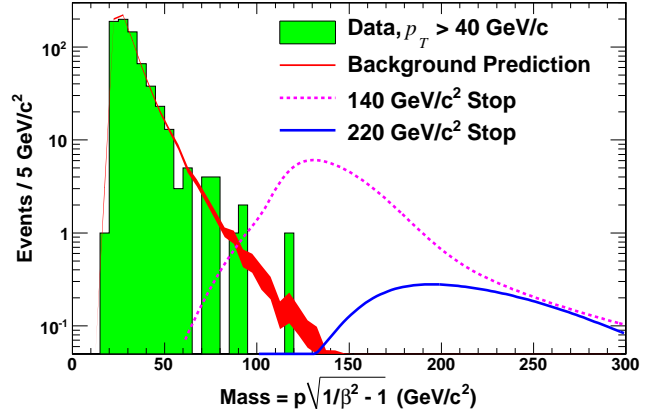}}
\caption{The mass of the CHAMP candidates as measured using the time-of-flight system. Taken from~\protect{\cite{CHAMPSearch}}.}
\label{fig: CHAMPEvents}
\end{figure}

\begin{figure}[ht]
\figurehelper{\includegraphics[width=8.66cm, height=6cm]{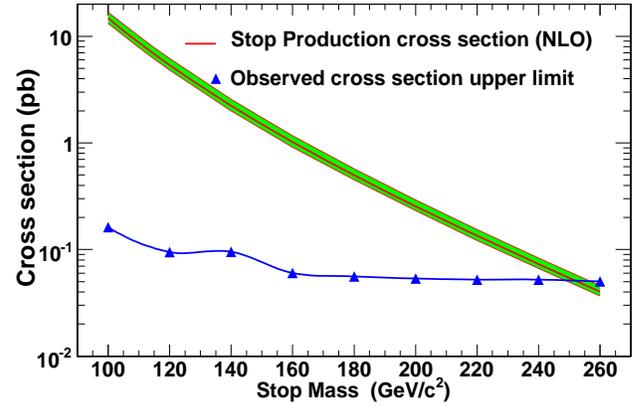}}
\caption{The 95\% confidence level cross section upper limits as a function of CHAMP mass in stop scenarios. Taken from~\protect{\cite{CHAMPSearch}}.}
\label{fig: CHAMPLimits}
\end{figure}

There is a new emphasis in the theory community about the role of long-lived sparticles in the early universe that decay into the dark matter particles we observe today~\cite{CHAMPTheory}. These particles can be identified in the detector because they are heavy, charged, and weakly interacting. These properties make them interact in the detector in ways that make them look like muons that are just traveling significantly slower than the speed of light. To search for these particles, timing techniques are used to measure the time of arrival for muon-like particles that arrive later than ``expected''~\cite{CHAMPSearch}. The results from 1.0~fb$^{-1}$,
converted to the ``measured mass'' are shown in Figure~\ref{fig: CHAMPEvents}, along with SM expectations. There is no evidence for new physics. The 95\% confidence level cross section upper limits where the CHAMP is a stop are shown in Figure~\ref{fig: CHAMPLimits}; similar results are expected when the limits are reinterpreted in stau and chargino scenarios. Comparable results from D\O~can be found in Ref.~\cite{D0CHAMP}.

\section{Gauge Mediated SUSY Breaking Models}\label{gmsb}

\mysect{$ee\gamma \gamma$ Candidate Event}

Gauge Mediated Supersymmetry Breaking (GMSB) models typically have the same mass relationships as in mSUGRA except for the \grav, the SUSY partner of the as-yet unobserved graviton. In mSUGRA models, the graviton is expected to have a mass at the GUT scale, but in GMSB models the \grav~is expected to have a mass of a keV, making it the LSP. Thus, the decay \none $\rightarrow \gamma$\grav~provides a warm dark matter candidate consistent with astronomical observations and models of inflation~\cite{gmsbtheory}. It has the further advantage that it provides a more natural solution for FCNC problems than does mSUGRA.
% These models also currently provide the only viable non-minimal model that can explain the $ee\gamma \gamma$\met~candidate event observed at CDF in Run~I~\cite{eeggmet}.

In typical models that take into account cosmological constraints~\cite{sps8} the \none~is favored to have a lifetime on the order of nanoseconds. This would make them stable on the timescale of the early universe, before they decayed into a \grav. Since limits on GMSB models exclude above $m_{\widetilde{\chi}_1^0} \leq 100$~GeV/$c^2$~\cite{FirstIIggMet}, this predicts that squarks and gluinos are too massive to be produced at the Tevatron. So, as in mSUGRA, gaugino pair production dominates. Each gaugino typically produces a \none, and other high $E_T$, light particles; the \none~decays via $\gamma \widetilde{G}$. The \grav, like the \none~in mSUGRA models, leaves the detector and can give significant \met. The lifetime and mass of the \none~dictate the different final states~\cite{tobwag}. Specifically, if both \none~decay in the detector, a final state of $\gamma \gamma$ + \met~+ $X$ is favored for a short lifetime ($t_{\widetilde{\chi}^0_1} \leq 1$~ns). For intermediate lifetimes, $1 \leq t_{\widetilde{\chi}_1^0} \leq 50$~ns, the $\gamma$ + \met~+ $X$ final state may be detectable if one \none~decays in the detector while the other leaves the detector without decaying or interacting. In this scenario, the time of arrival of the photon at the calorimeter can be the signal of a ``long'' lifetime \none. For large lifetimes, both neutralinos can leave the detector and are indistinguishable from mSUGRA scenarios.  The different lifetime scenarios are considered separately.

\subsection{Short-Lived}\label{ggmet}

\mysect{Large Numbered $\gamma \gamma$ + \met~Searches}

\begin{figure}[htb]
%\figurehelper{\includegraphics[width=8.66cm]{GMSBLifetimeLimits}}
\figurehelper{\includegraphics[width=8.66cm]{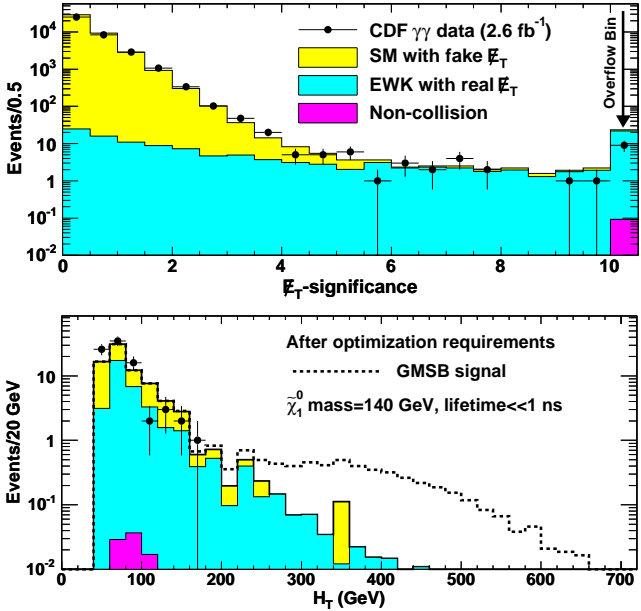}}
\caption{(Top) The \met~significance distribution in inclusive $\gamma \gamma$ events. (Bottom) The $H_T$ distribution of $\gamma \gamma$ + \met~events in the search for low lifetime \none~in GMSB events. Taken from~\protect{\cite{GMSBgammagammamet}}.}
\label{fig: GMSBLifetimeLimits}
\end{figure}

\begin{figure}[htb]
\figurehelper{\includegraphics[width=8.3cm]{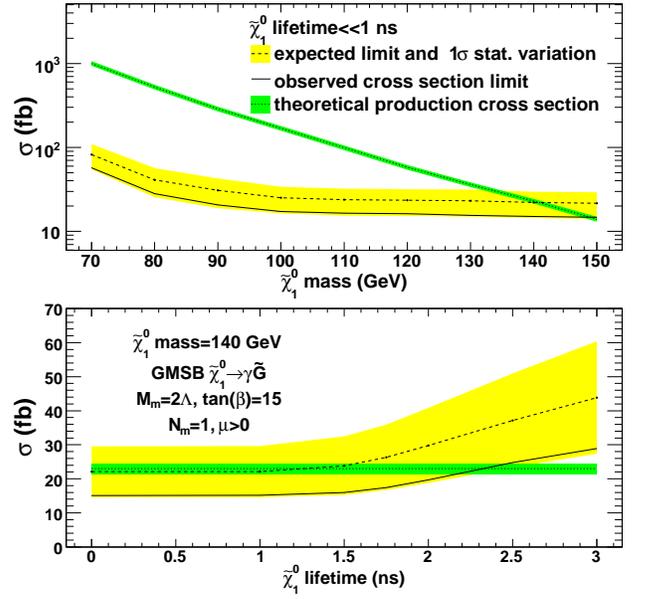}}
%\figurehelper{\includegraphics[width=8.66cm]{SigVsLife}}
%\figurehelper{\includegraphics[width=8.66cm]{SigVsMass}}
\caption{Cross section limits on sparticle production in GMSB models as a function of \none~mass for $\tau_{\widetilde{\chi}^0_1} = 0$ and as a function of $\tau_{\widetilde{\chi}^0_1}$ at $m_{\widetilde{\chi}_1^0}$ = 140~GeV/$c^2$. Taken from~\protect{\cite{GMSBgammagammamet}}.}
\label{fig: SigsVs}
\end{figure}

For short lived \none~the dominant search channel is $\gamma \gamma$ + \met. CDF has recently developed a new tool to assess the significance of the \met~measurement~\cite{ggxprd}. This allows for a straightforward separation of QCD backgrounds with no intrinsic \met~from those with real \met~such as EWK sources. Also a new photon timing device is used to remove non-collision sources of background~\cite{emtim}. A signature-based search in 2.6~fb$^{-1}$ of data for $\gamma \gamma$ events with significant \met~is shown in Figure~\ref{fig: GMSBLifetimeLimits}~\cite{GMSBgammagammamet, ggxprd}. To optimize the $\gamma \gamma$ + \met~search for $\tau_{\widetilde{\chi}_1^0} \ll$  1~ns, the GMSB model analysis requires both significant \met~and large $H_T$ to identify the extra particles produced in the decay of the gauginos~\cite{GMSBgammagammamet}. After optimization, the results are shown in the bottom of Figure~\ref{fig: GMSBLifetimeLimits}. Since there is no evidence for new physics, cross section limits are set on sparticle production as a function of the \none~mass in Figure~\ref{fig: SigsVs}. Comparable results from D\O~that assume $\tau_{\widetilde{\chi}_1^0}\ll$~1~ns can be found at~\cite{D0ShortLivedGMSB}. Of particular note is that the null results can also be interpreted as limits on production as a function of \none~lifetime as shown in the bottom of Figure~\ref{fig: SigsVs}. 

\subsection{Long-Lived}\label{delayed}

\mysect{All Neutralino Lifetime Searches}

\begin{figure} [h]
\figurehelper{\includegraphics[width=8.66cm]{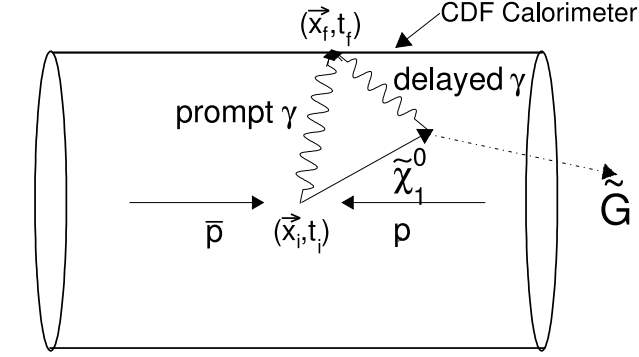}}
\figurehelper{\includegraphics[width=8.66cm]{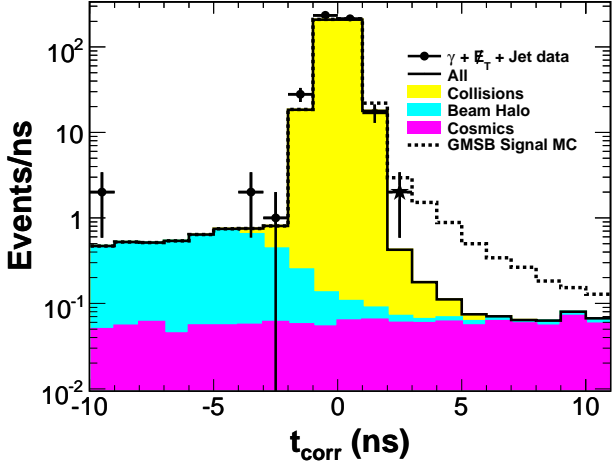}}
\caption{(Top) A diagram of the how photons from long lived decays of \none~$\rightarrow \gamma$\grav~arrive at the detector with delayed times. (Bottom) The corrected timing distribution for the delayed photon analysis in the $\gamma$ + jet + \met~search for long lifetime \none~GMSB models. Taken from~\protect{\cite{HeavyPartToPhot}}.}
\label{fig: DelayedPhotonGMSB}
\end{figure}

\begin{figure} 
\figurehelper{\includegraphics[width=8.66cm]{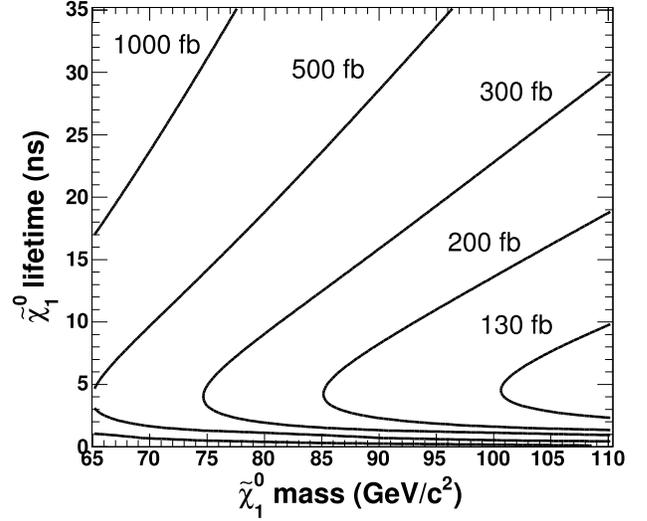}}
\caption{Contours of constant 95\% confidence level cross section upper limits on GMSB SUSY production from the $\gamma$ + jet + \met~final state search for long-lived \none 's. Taken from~\protect{\cite{HeavyPartToPhot}}.}
\label{fig: Contours}
\end{figure}

\begin{figure}
%\figurehelper{\includegraphics[width=8.66cm]{GMSBLifetimeVsMass}}
\figurehelper{\includegraphics[width=8.66cm]{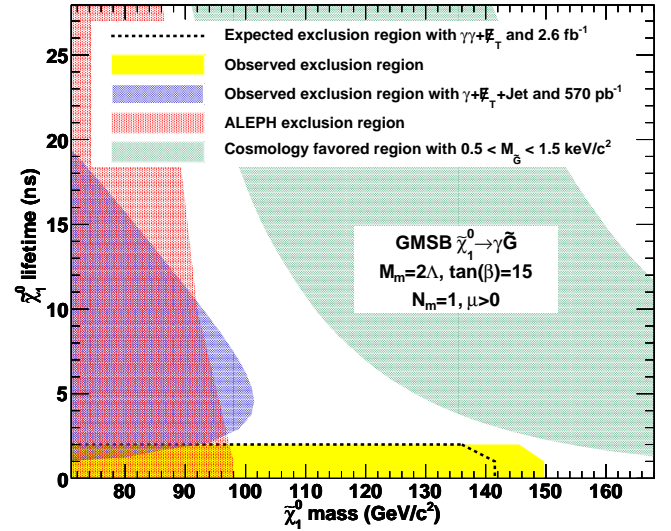}}
\caption{The combined GMSB exclusion regions in the lifetime vs. mass plane of the \none. Taken from~\protect{\cite{GMSBgammagammamet}}.}
\label{fig: GMSBLifetimeVsMass}
\end{figure}

Cosmological constraints on GMSB favor keV \grav~masses and nanosecond \none~lifetimes for \none~masses above 100~GeV/$c^2$. At CDF a photon timing system is used to measure the time of arrival of photons in the detector~\cite{emtim}, Figure~\ref{fig: DelayedPhotonGMSB} (Top) shows the technique for distinguishing prompt photon production from ``delayed'' photons from long lifetime \none~that travel macroscopic distances then decay in the detector. The search for long-lived neutralinos requires a photon, large \met~and a jet from the gaugino decays~\cite{HeavyPartToPhot} in 570~pb$^{-1}$ of data. Figure~\ref{fig: DelayedPhotonGMSB} (Bottom) shows the photon time of arrival corrected for the collision time and the time of flight for the sample. There is no evidence for new physics. Figure~\ref{fig: Contours} shows the 95\% confidence level cross section upper limits in the lifetime vs. mass plane. Figure~\ref{fig: GMSBLifetimeVsMass} shows the combined exclusion region from both analyses. Projections indicate that with 10 fb$^{-1}$ the search sensitivity should be well into the cosmology favored region.

\newcommand{\oldother}{
\section{Other Possibilities}\label{other}

\mysect{Lots of Other Possibilities}

There are two other SUSY scenarios worth mentioning here. The first are CHAMPS, which are CHArged Massive quasi-stable ParticleS~\cite{CHAMPSearch}. This model is like GMSB in that the lightest and most abundant sparticle in the early universe is different than it is today, but like mSUGRA in that the LSP is the \none. A second model is $R$-parity violating SUSY. In this model, nature might be supersymmetric, but may also have nothing to do with dark matter. As previously mentioned, it is still worth looking into this possibility, but it's more difficult to narrow down the possibilities so as to be able to focus the searches for these new particles.

\subsection{R-Parity Violating SUSY}\label{rpv}

\mysect{R-Parity Violating SUSY}

One advantage of RPV SUSY is that without $R$-Parity conservation, single sparticle production is allowed so the production cross sections can be significantly large compared to pair production for the same sparticle mass. Decays, however, depend critically on the couplings and there are now no final state SUSY particles, nor a dark matter candidate. In the past, there have been searches for mass resonances for single neutralino production that decay via $e \mu$ at CDF~\cite{HighMassRestoLept} and D\O~\cite{D0RPV}. Recent results with 1.0~fb$^{-1}$ of data expand from $e\mu$, to search for stau sneutrinos that decay into the $e\tau$ and $\mu \tau$ final states using powerful new tau identification tools. Backgrounds are dominated by EWK and W + Jet backgrounds where the jet fakes a hadronic tau decay. The data in all three modes are shown in Figure~\ref{fig: RPV}. Since there is no evidence for new physics, cross section limits are set, as shown in Figure~\ref{fig: SneutrinoDecayLimits} in three different coupling scenarios.

\begin{figure}
\figurehelper{\includegraphics[width=8.66cm, height=4cm]{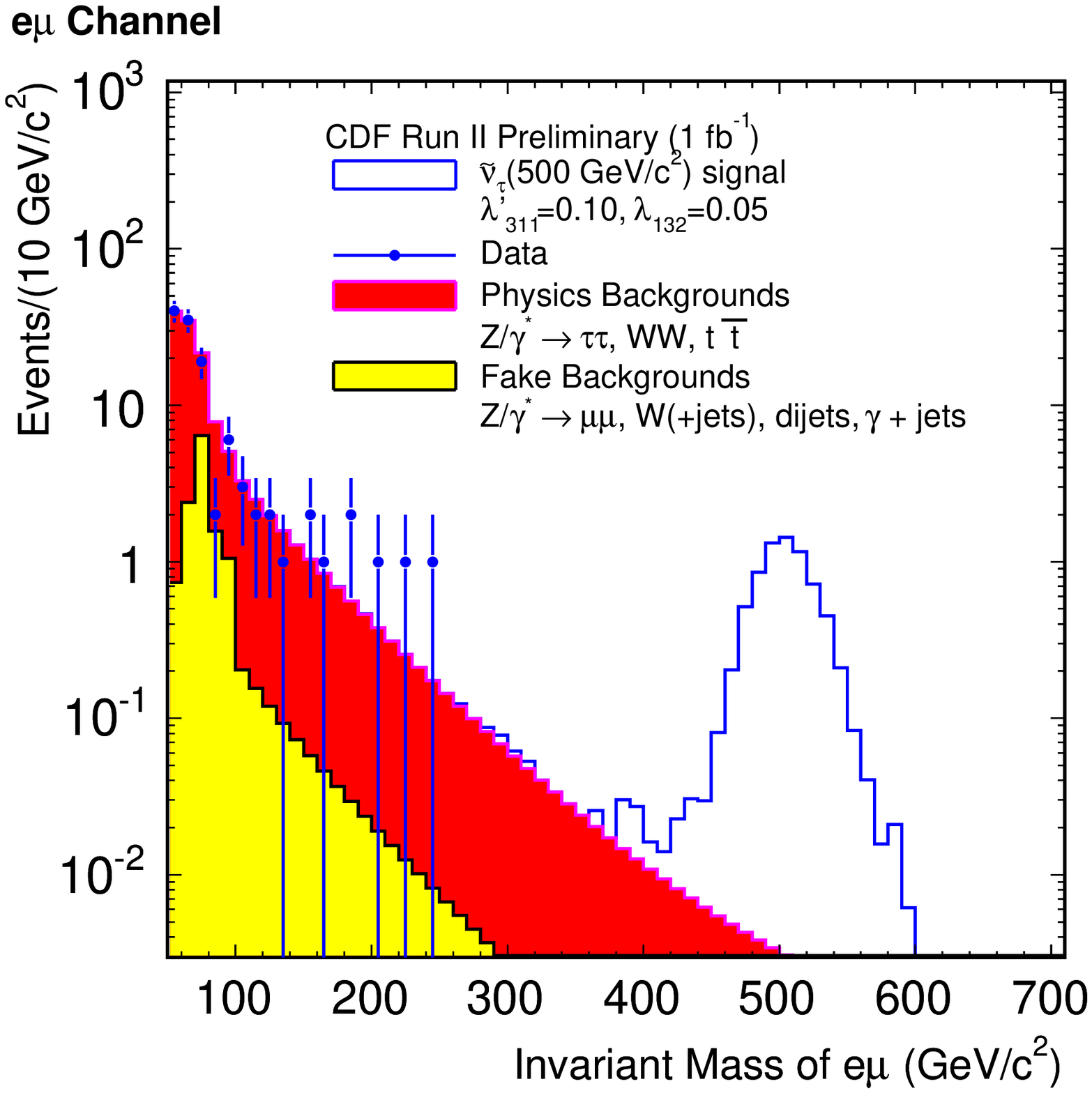}}
\figurehelper{\includegraphics[width=8.66cm, height=4cm]{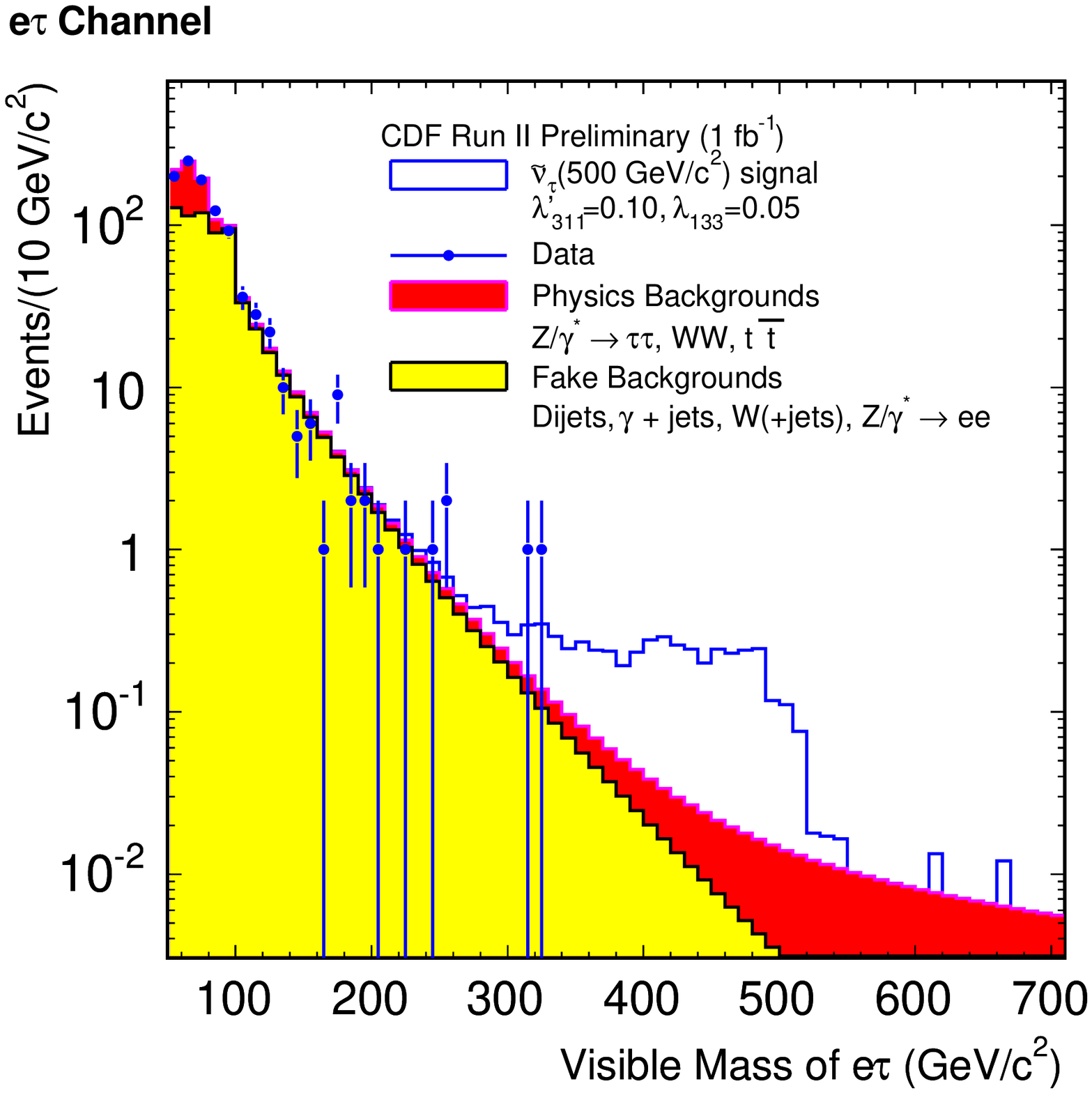}}
\figurehelper{\includegraphics[width=8.66cm, height=4cm]{RPVmtChan}}
\caption{The invariant mass distribution in the search for sneutrino production in $R$-parity violating SUSY in (Top) the $e\mu$ channel, (Middle) the $e \tau$ channel, and (Bottom) the $\mu \tau$ channel. Taken from~\protect{\cite{HighMassRestoLept}}.}
\label{fig: RPV}
\end{figure}

\begin{figure}
\figurehelper{\includegraphics[width=8.66cm, height=12cm]{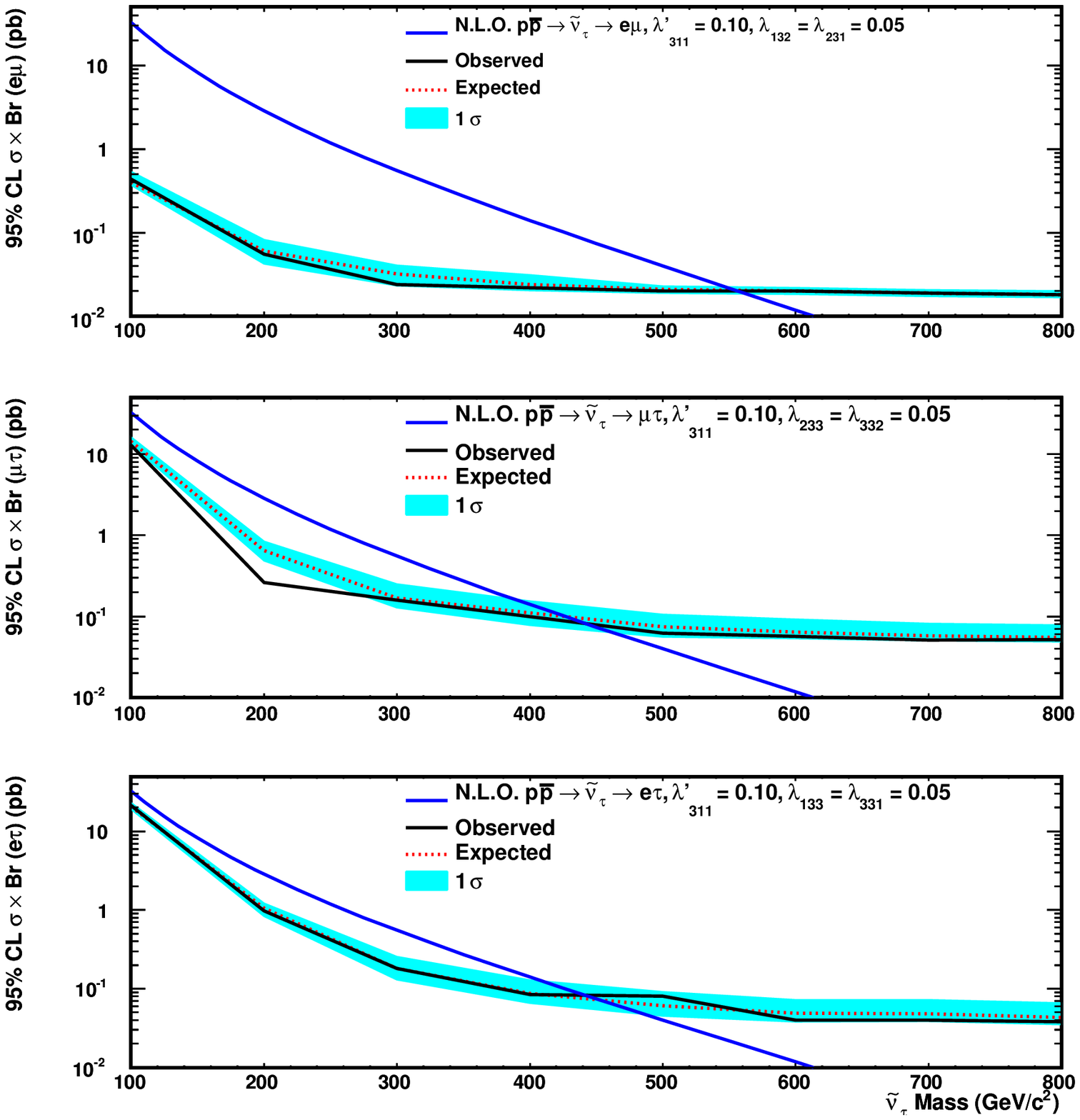}}
\caption{The 95\% confidence level cross section limits on production and $R$-parity violating decay of single sneutrinos in (Top) the $e\mu$, (Middle) the $e \tau$, and (Bottom) $\mu \tau$ final states. Taken from~\protect{\cite{HighMassRestoLept}}.}
\label{fig: SneutrinoDecayLimits}
\end{figure}
}

\section{Conclusions}

\mysect{Conclusions}

The CDF experiment has performed a broad and deep set of cosmology inspired searches for SUSY. Unfortunately, it has found no evidence for new physics. As of mid-2009, the Tevatron is still delivering data at unprecedented levels and the detectors are functioning beautifully. With data taking again started in 2009, the future is bright for discovery at the Tevatron. Until the LHC starts running, it is still leading the search for SUSY.

\section{Acknowledgments}

\mysect{Thanks}

First and foremost, the author would like to thank Sean Yeager of the Texas A\&M/NSF REU Program for his help in the preparation of this document. He also would like to thank David Rahmani for his help in producing Figure~\ref{fig: FeynmanDecays}. Finally, he would like to thank Richard Arnowitt, Monica D'Onofrio, Sourabh Dube, Bhaskar Dutta, Andrew Ivanov, Teruki Kamon, Sasha Pronko, Alexei Safonov, and Sunil Somalwar for their help and feedback on this document.

\appendix
\section{Notes on the LHC Startup}
\MyAside

\end{document}

%% file: definitions.tex
% definitions for CDF notes: include with \input{definitions}

\newcommand{\mysect}[1]{}%{\addtocounter{myctr}{1} \noindent \underline{Paragraph {\arabic{myctr}}: #1}}
\newcommand{\refhelper}[1]{}
\newcommand{\figurehelper}[1]{#1}

%\def\gsim{\mathrel{\rlap{\lower4pt\hbox{\hskip1pt$\sim$}}
%    \raise1pt\hbox{$>$}}} 

% use ${X_1} to use math definitions for math and text mode !!!
\newcommand{\munit}{$\mathrm{\frac{GeV}{c^2}}$}
\newcommand{\punit}{$\mathrm{GeV}/c$}
\newcommand{\mgunit}{$\mathrm{eV}/c^2$}
\newcommand{\mkunit}{$\mathrm{keV}/c^2$}
\newcommand{\lumunit}{{$\rm cm^{-2}\ s^{-1}$}}
\newcommand{\dedxunit}{$\frac{\rm keV}{\rm cm}$}

\newcommand{\et}{{E\!_T}}
\newcommand{\met}{${E\!\!\!\!/_T}$}
\newcommand{\mett}{\mbox{${E\!\!\!\!/_T}$}}
\newcommand{\metx}{\mbox{${E\!\!\!\!/_{T}~\!\!\!\!^{x}}$}}  % S.Lee   (Mar  2003)
\newcommand{\mety}{\mbox{${E\!\!\!\!/_{T}~\!\!\!\!^{y}}$}}  % S.Lee   (Mar  2003)
\newcommand{\mmetx}{\mbox{Mean $(E\!\!\!\!/_{T}~\!\!\!\!^{x})$}}
\newcommand{\mmety}{\mbox{Mean $(E\!\!\!\!/_{T}~\!\!\!\!^{y})$}}
\newcommand{\mpt}{${p\!\!\!\!/_T}$}
\newcommand{\summet}{\mbox{$\sigma(E\!\!\!\!/_T)$}}
\newcommand{\summetx}{\mbox{$\sigma(E\!\!\!\!/_{T}~\!\!\!\!^{x})$}}
\newcommand{\summety}{\mbox{$\sigma(E\!\!\!\!/_{T}~\!\!\!\!^{y})$}}
\newcommand{\sumet}{${\Sigma {\rm E_{T}}}$}
\newcommand{\sumpt}{${\Sigma {p_{T}}}$}
\newcommand{\sumpttrk}{${\Sigma {p_{\rm T}^{\mathrm{trk}}}}$}
\newcommand{\dedx}{${\frac{\mathrm{d}E}{\mathrm{d}x}}$}

\newcommand{\chisq}{${\chi^{2}}$}
\newcommand{\mN}{${m_{%\widetilde
{\none}}}$}
\newcommand{\tauN}{${\tau_{%\widetilde
{\none}}}$}

\newcommand{\tcorr}{${t_{\mathrm{corr}}}$}
\newcommand{\tarr}{${t^{\gamma}_{\mathrm{corr}}}$}

\newcommand{\etc}{\mbox{$\Sigma {\rm E_{T}^{Corrected}}$}}
\newcommand{\setc}{\mbox{$\sqrt{\Sigma {\rm E_{T}^{Correc
ted}}}$}}
\newcommand{\Zee}{\mbox{$Z \rightarrow ee$}}
\newcommand{\Wjets}{\mbox{$W+\mathrm{jets}$}}
\newcommand{\Zmumu}{\mbox{$Z \rightarrow \mu\mu$}}
\newcommand{\Wenu}{${W \rightarrow e\nu}$}
\newcommand{\Wenuj}{${W \rightarrow e\nu+\mathrm{jets}}$}
\newcommand{\Wnotrack}{{\tt W\_NOTRACK}}
\newcommand{\Metpem}{MET\_PEM}
\newcommand{\Zeg}{$Z \rightarrow e\gamma$}

\newcommand{\gmetjets}{${\gamma+\mathrm{jet}+\met}$}
\newcommand{\gmetjet}{${\gamma+\mathrm{jet}+\met}$}
\newcommand{\ggmet}{${\gamma\gamma+\met}$}
\newcommand{\gmet}{${\gamma+\met}$}
\newcommand{\eeggmet}{${ee\gamma\gamma\met}$}

\newcommand{\gt}{\mbox{$>$}}
\newcommand{\lt}{\mbox{$<$}}

\newcommand{\gev} {\rm \,GeV}
\newcommand{\gevt} {$\rm \,GeV^2$}
\newcommand{\ipb}{\mbox{${\rm pb}^{-1}$}}
\newcommand{\dphi}{${\Delta\phi(\met,\rm{jet})}$}
\newcommand{\dphicosm}{${\Delta\phi(\mu\mathrm{-stub},\gamma)}$}

\newcommand{\pt}{${p_{_T}}$}
\newcommand{\tzero}{${t_{0}}$}
\newcommand{\zzero}{${z_{0}}$}
\newcommand{\etg}{${E_{T}^{\gamma}}$}
\newcommand{\etjet}{${E_{T}^{\mathrm{jet}}}$}
\newcommand{\etajet}{${\eta^{\mathrm{jet}}}$}

\newcommand{\chargino}{$\widetilde{\chi}_{1}^{\pm}$}
\newcommand{\ncl}{$\rm N_{95\%}$}
\newcommand{\sigexp}{${\sigma^{\mathrm{exp}}_{95}}$}
\newcommand{\sigobs}{${\sigma^{\mathrm{obs}}_{95}}$}
\newcommand{\sigmc}{${\sigma_{\mathrm{Signal\ MC}}}$}
\newcommand{\sigprod}{${\sigma_{\mathrm{prod}}}$}

\newcommand{\grav}{${\widetilde{G}}$}
\newcommand{\none}{${\widetilde{\chi}_1^0}$}
\newcommand{\ntwo}{${\widetilde{\chi}_2^0}$}
\newcommand{\nthree}{${\widetilde{\chi}_3^0}$}
\newcommand{\nfour}{${\widetilde{\chi}_4^0}$}
\newcommand{\conep}{${\widetilde{\chi}_1^{+}}$}
\newcommand{\conem}{${\widetilde{\chi}_1^{-}}$}
\newcommand{\conepm}{${\widetilde{\chi}_1^{\pm}}$}
\newcommand{\conemp}{${\widetilde{\chi}_1^{\mp}}$}
\newcommand{\ctwopm}{${\widetilde{\chi}_2^{\pm}}$}
\newcommand{\stau}{${\widetilde{\tau}}$}
\newcommand{\decay}{\mbox{$p\bar{p}\rightarrow X\rightarrow \none\none$}}
\newcommand{\nonetogG}{${\none\rightarrow\gamma\grav}$}
\newcommand{\Mm}{M_{m}}
\newcommand{\etal}{{\em et al.}}

\newcommand{\ag}{${\alpha}$}
\newcommand{\bg}{${\beta}$}
\newcommand{\dg}{${^{\circ}}$}%  % degree symbol:  °
\newcommand{\rphi}{$(r,\phi)$}
\newcommand{\rz}{$r,z$}

\newcommand{\mum}{${\mu{\rm m}}$}
\newcommand{\mus}{${\mu{\rm s}}$}

\def\bi{\begin{itemize}}
\def\ei{\end{itemize}}
\def\bc{\begin{center}}
\def\ec{\end{center}}
\def\and{\/\mbox{and}}
\def\egg{$e\gamma\gamma$}
\def\egga{$e\gamma_{13}\gamma_{13}$}
\def\eggb{$e\gamma_{25}\gamma_{25}$}
\def\eg{$e\gamma$}
\def\g{$\gamma$}
\def\mgg{$\mu\gamma\gamma$}
\def\mgga{$\mu\gamma_{13}\gamma_{13}$}
\def\mggb{$\mu\gamma_{25}\gamma_{25}$}
\def\mga{$\mu\gamma_{13}$}
\def\mgb{$\mu\gamma_{25}$}
\def\mg{$\mu\gamma$}
\def\m{$\mu$}
\def\etog{$e\rightarrow \gamma$}
\def\eetoeg{$ee\rightarrow e\gamma$}
\def\etoj{$e\rightarrow jet$}

\def\bvec#1{\vec{\bf #1}}
\newcommand{\rd}{{\rm d}}
\def\invpb{\mbox{pb$^{-1}$}}
\def\invfb{\mbox{fb$^{-1}$}}
\def\invnb{\mbox{nb$^{-1}$}}

\def\pythia{{\sc pythia}}
\def\prospino{{\sc prospino2}}

\def\Journal#1#2#3#4{{#1}\textbf{~#2}, #4 (#3)}
\def\PrePrint#1{\mbox{hep-ph/#1}}
\def\PRL{\rm Phys. Rev. Lett.}
\def\PRD{{\rm Phys. Rev.} D}

\newcommand{\pbar}{${\overline{p}}$}
\newcommand{\ppbar}{${p\overline{p}}$}
\newcommand{\cdf}{CDF~II}
\newcommand{\vev}{\text{\it vev}}
\newcommand{\sign}{${\mathrm{sign}}$}

\newcommand{\tevE}{${\sqrt{s} = 1.96~\TeV}$}
\newcommand{\SORaby}{${\mathrm{MSO}_{10}\mathrm{SM}}$}